\documentclass[%
reprint,
aps,
prb,
floatfix,
]{revtex4-2}

\usepackage{hyperref}
\usepackage{dcolumn}
\usepackage{bm}
\usepackage{amsmath}
\usepackage{natbib}
\usepackage{booktabs}
\DeclareMathAlphabet{\mathbit}{OT1}{cmr}{bx}{it}

\usepackage{docmute}
\usepackage{color}
\usepackage{ulem}
\usepackage{multirow}

\usepackage{amsmath}
\usepackage{amsfonts}
\usepackage{amssymb}
\usepackage{graphicx}
\usepackage{braket}

\begin{document}

%

\title{Topological superconductivity in helical crystals}

\author{Soma Yoshida$^{1}$, Keiji Yada$^{1}$, Yukio Tanaka$^{1}$, Takehito Yokoyama$^{2}$}
\affiliation{%
  $^1$~Department of Applied Physics, Nagoya University, Nagoya 464--8603, Japan\\
  $^2$~Department of Physics, Tokyo Institute of Technology, Tokyo 152--8551, Japan
}%

\begin{abstract}  
We study superconductivity and surface Andreev bound states in helical crystals. We consider the interlayer pairings along the helical hopping and investigate the surface local density of states on the (001) and zigzag surfaces for all the possible irreducible representations.
There are three and four irreducible representations exhibiting the zero energy peaks in the local density of states at the (001) and zigzag surfaces of helical lattices, respectively.
By calculating the one dimensional winging number, we show that these appearances of the zero energy peaks stem from the surface Andreev bound states. 
\end{abstract}

\pacs{pacs}
\maketitle

\thispagestyle{empty}

\section{INTRODUCTION}

The symmetries of pair potentials are related to those of the underlying crystals\cite{sigrist_RMP_1991}. 
For example, in the system with inversion symmetry, the symmetry of the pair potential is classified into the even-parity spin-singlet or odd-parity spin triplet states.
In the transition of an unconventional superconductor (SC), one or more symmetries are broken in addition to U(1) symmetry breaking in BCS SCs.
Allowed pair potentials in the underlying crystal lattice structure are classified by the irreducible representations of the point group of the crystal lattice. 
The symmetry of the pair potential has been extensively studied in several SCs: cuprate, UTe$_2$, and SrRuO$_4$\cite{Tsuei_RMP_2000, Kallin_RPP_2016, Jiao_Nature_2020, Hirschfeld_RPP_2011, Aoki_JP_2022}.

Helical crystals, realized in materials such as tellurium\cite{Reitz_PR_1957,Caldwell_PR_1959,Hulin_JPCS_1966,Laude_PRL_1971,Nakayama_PRB_2017,Tsirkin_PRB_2018,sakano_PRL_2020}, have the right or left handedness.
The superconductivity in helical crystals has been found in, e.g., NbRh$_2$B$_2$ and TaRh$_2$B$_2$\cite{Carnicom_SA_2018,Mayoh_PRB_2018,Matano_PRB_2021}.
As a result of the helical crystal structures, current-induced orbital and spin magnetizations in helical crystals have been theoretically proposed\cite{Yoda_SR_2015,Yoda_NL_2018}, and chirality-induced spin selectivity (CISS)\cite{Ray_Science_1999,Gohler_Science_2011,Naaman_ARPC_2015,Michaeli_CSR_2016,Naaman_JPCL_2020,Waldeck_APLM_2021,Evers_AM_2022} has been detected in helical crystals\cite{Inui_PRL_2020,Shiota_PRL_2021,Shishido_APL_2021,Nakajima_nature_2023}.
These effects inducing the magnetization by the electric current are useful for the application to spintronics.

The effect of helical molecules chemisorbed on the conventional SC has been reported in the recent experiments\cite{Alpern_NJP_2016, Alpern_NL_2019, Alpern_PRM_2021}.
Conductance spectra are observed through the helical molecules in the spin-singlet s-wave SC (Nb) by the STS and STM measurements.
Interestingly, they show zero bias conductance peaks.
This result is against the fact that a zero-bias conductance peak is not exhibited on the surface of $s$-wave SCs because the anisotropy of the gap function such as $p$-wave or $d$-wave SCs is necessary to generate the zero energy bound states on the surface\cite{andreev_SPJ_1964,Buchholtz_PRB_1981,Hara_PTP_1986,Hu_PRL_1994,Tanaka_PRL_1995,Kashiwaya_RPP_2000,Lofwander_SST_2001,Asano_PRB_2004}.
Thus, it is suggested that unconventional superconductivity is proximity induced in the helical molecules.
The experimental result performed in the helical molecules intercalated into a layered SC suggests that the molecular chirality induces the unconventional/topological SC\cite{Zhong_arXiv_2023}.
However, the mechanism of this effect of the helical molecules has not been established yet, while this zero bias conductance peak structure implies the possibility of novel effect of helical structures.

\begin{figure}[tbp]
	\centering
	\includegraphics[width=0.9\columnwidth]{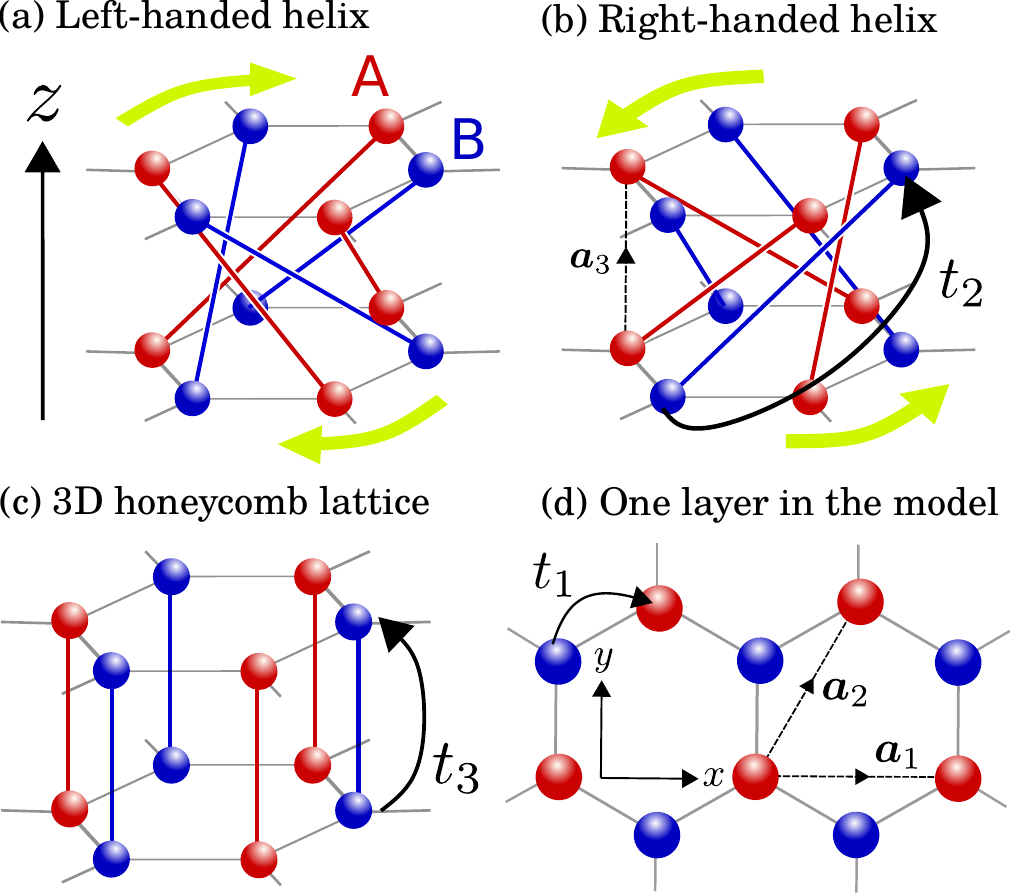}
	\caption{Helical lattice of the (a) left-handed helix and (b) right-handed helix, where $\bm{a}_3$ is a primitive lattice vector along the $z$ axis. 
	(c) Three dimensional (3D) honeycomb lattice is shown as a reference. 
	(d) One layer of the helical and honeycomb lattice, where $\bm{a}_1$ and $\bm{a}_2$ denote the primitive lattice vectors on the $xy$-plane.
	The A and B sites are marked by red and blue balls, respectively. Red and blue bonds in (a), (b) and (c) show the interlayer hoppings between A and B sites, respectively. 
	The hopping amplitude $t_1$ is the nearest neighbor hopping in the $xy$ plane, and} $t_2$ and $t_3$ are the interlayer hopping amplitudes along bonds in the helical and 3D honeycomb lattices, respectively.
	\label{fig:lattice}
\end{figure}

It is known that the dispersionless Andreev bound states (ABSs) are manifested as zero bias conductance peaks on the surface of the unconventional SCs.
The presence of the zero energy flat-band ABSs on the surface is characterized by the topological number (winding number) defined in the bulk system\cite{sato_PRB_2011}.
A SC with non-trivial winding number is identified with the topological SC\cite{Qi_RMP_2011,Leijnse_SST_2012,Beenakker_ARCMP_2013,Schnyder_JP_2015,tanaka_JPSJ_2012,chiu_RMP_2016,Sato_JPSJ_2016,Sato_RPP_2017,Marra_JAP_2022}, and the bound states protected by the winding number are robust against any perturbations as long as the system remains the symmetry to define the topological number.
Thus, it is interesting to investigate the ABSs and winding number in the system with helical structures to clarify the symmetry of the pairing in the helical systems.

\begin{figure}[tbp]
	\centering
	\includegraphics[width=0.9\columnwidth]{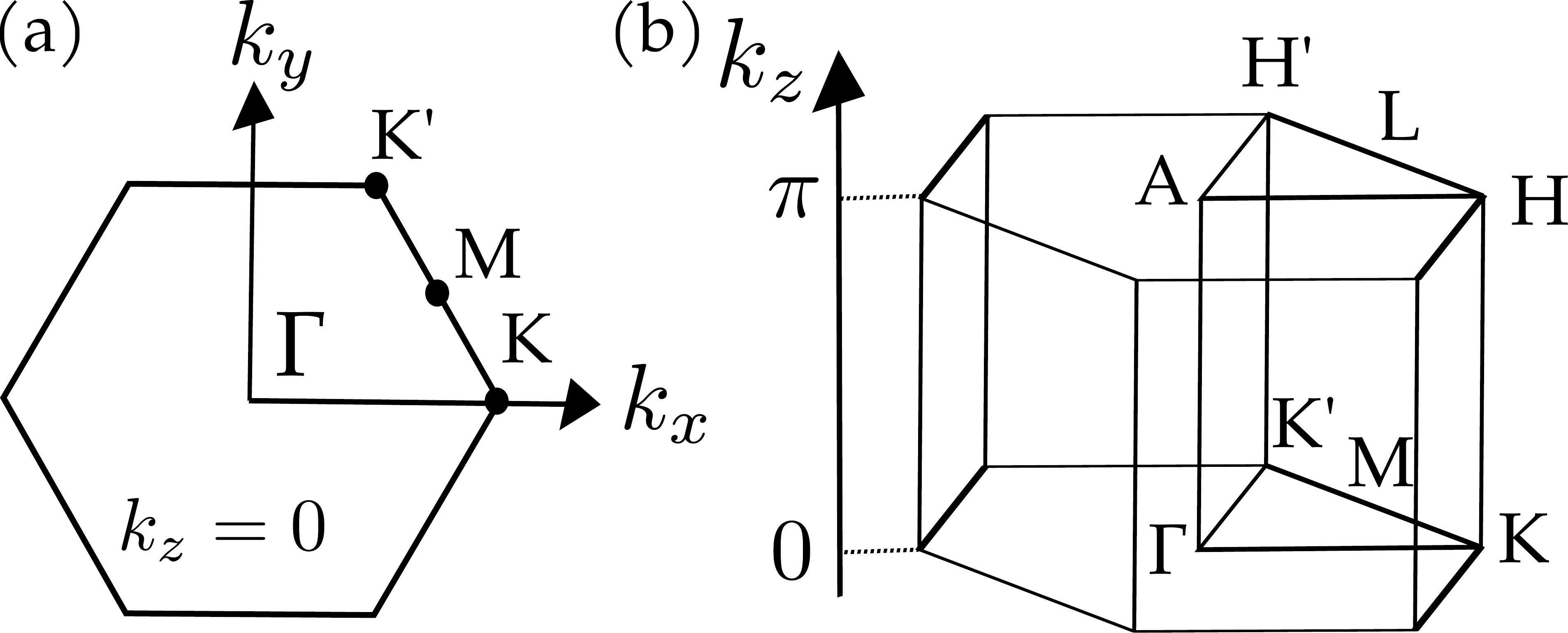}
	\caption{(a) High symmetry points on $k_z=0$ plane in the Brillouin zone. 
	(b) High symmetry points in the half range of the Brillouin zone. 
	The origin of the Brillouin zone corresponds to $\Gamma$ point.}
	\label{fig:hsp}
\end{figure}

\begin{figure}[tbp]
	\centering
	\includegraphics[width=0.9\columnwidth]{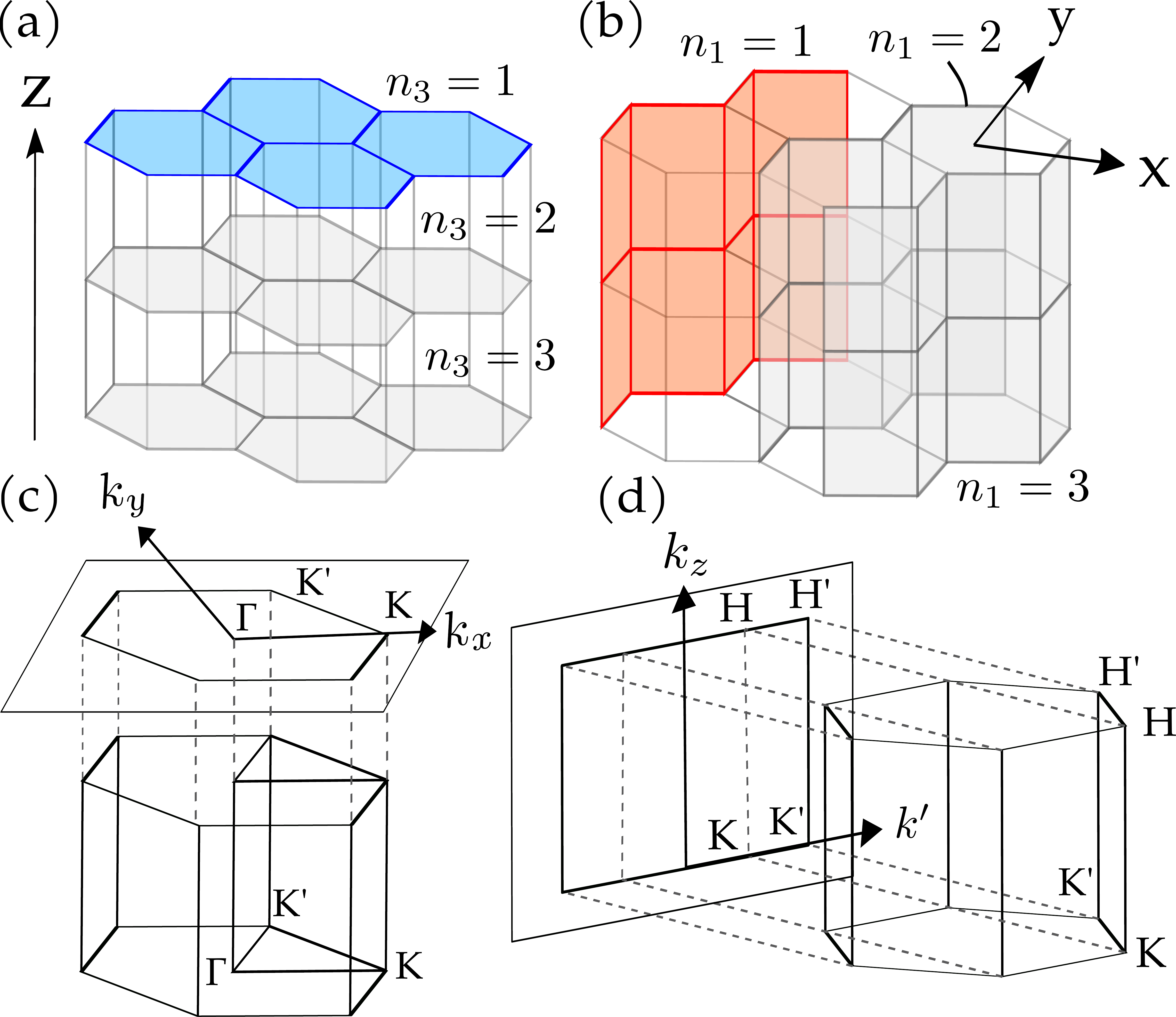}
	\caption{Semi-infinite models with (a) (001) and (b) zigzag surface.
		The blue and red planes show the (001) and zigzag surface, respectively.
		The layers parallel to the (001) (zigzag) surface are labeled by $n_3$ ($n_1$), 
		where $n_3=1$ ($n_1=1$) layer corresponds to the surface.
		Surface Brillouin zone projected to the (c) (001) and (d) zigzag surface. 
	        The axis $k^\prime$ is aligned parallel to the zigzag surface on the $k_x k_y$-plane. 
	}
	\label{fig:surface}
\end{figure}

\begin{table*}[hbtp]
\caption{
	Irreducible representations (Irreps) and basis functions $\phi_{\mu}^{IR}(\bm{k})$ of the pair potentials for interlayer pairing in the honeycomb and helical lattices, where $\mu$ indicates the sub-lattice degree of freedom.
	$D_{6h}$ and $D_6$ represent the point groups (PG) in the honeycomb and helical lattices, respectively.
	The inter-site components are zero because we focus on the interlayer pairings.
	Node structures are obtained at $t_2/t_1=0.1$ or $t_3/t_1=0.1$ and $\Delta_0/t_1=0.1$. 
	}
  \centering
  \begin{tabular}{cccccc}
	  \hline PG & Irrep & Spin & Node & $\phi_A^{IR}(\bm{k})$ & $\phi_B^{IR}(\bm{k})$ \\
    \hline\hline
	$D_{6h}$ & $A_{1g}$ & singlet & point & ~~$\cos{k_z}$ ~~ & ~~$ \phi_A(\bm{k}) $~~ \\
		 & $A_{2u}$ & triplet & line  & ~~$\sin{k_z}$ ~~ & ~~$ \phi_A(\bm{k}) $~~ \\
		 & $B_{1u}$ & singlet &       & ~~$\cos{k_z}$ ~~ & ~~$-\phi_A(\bm{k}) $~~ \\
		 & $B_{2g}$ & triplet &       & ~~$\sin{k_z}$ ~~ & ~~$-\phi_A(\bm{k}) $~~ \\
	$D_{6 }$ & $A_1   $ & singlet & point & ~~$\cos(k_x+k_z)+\cos(k_x/2-\sqrt{3}k_y/2-k_z)+\cos(k_x/2+\sqrt{3}k_y/2-k_z)$ ~~ & ~~$ \phi_A(k_x,k_y,-k_z)$~~ \\
		 & $A_2   $ & triplet & line  & ~~$\sin(k_x+k_z)-\sin(k_x/2-\sqrt{3}k_y/2-k_z)-\sin(k_x/2+\sqrt{3}k_y/2-k_z)$ ~~ & ~~$-\phi_A(k_x,k_y,-k_z)$~~ \\
		 & $B_1   $ & singlet &       & ~~$\cos(k_x+k_z)+\cos(k_x/2-\sqrt{3}k_y/2-k_z)+\cos(k_x/2+\sqrt{3}k_y/2-k_z)$ ~~ & ~~$-\phi_A(k_x,k_y,-k_z)$~~ \\
		 & $B_2   $ & triplet &       & ~~$\sin(k_x+k_z)-\sin(k_x/2-\sqrt{3}k_y/2-k_z)-\sin(k_x/2+\sqrt{3}k_y/2-k_z)$ ~~ & ~~$ \phi_A(k_x,k_y,-k_z)$~~ \\
		 & \multirow{2}{*}{$E_1$} & \multirow{2}{*}{singlet} & \multirow{2}{*}{line} & ~~$2\cos(k_x+k_z)-\cos(k_x/2-\sqrt{3}k_y/2-k_z)-\cos(k_x/2+\sqrt{3}k_y/2-k_z)$~~ & ~~$-\phi_A(k_x,k_y,-k_z)$~~ \\
		 &                        &                          &                       & ~~$-\cos(k_x/2-\sqrt{3}k_y/2-k_z)+\cos(k_x/2+\sqrt{3}k_y/2-k_z)$              ~~ & ~~$-\phi_A(k_x,k_y,-k_z)$~~ \\
		 & \multirow{2}{*}{$E_1$} & \multirow{2}{*}{triplet} & \multirow{2}{*}{line} & ~~$-\sin(k_x/2-\sqrt{3}k_y/2-k_z)+\sin(k_x/2+\sqrt{3}k_y/2-k_z)$              ~~ & ~~$ \phi_A(k_x,k_y,-k_z)$~~ \\
		 &                        &                          &                       & ~~$2\sin(k_x+k_z)+\sin(k_x/2-\sqrt{3}k_y/2-k_z)+\sin(k_x/2+\sqrt{3}k_y/2-k_z)$~~ & ~~$ \phi_A(k_x,k_y,-k_z)$~~ \\
		 & \multirow{2}{*}{$E_2$} & \multirow{2}{*}{singlet} & \multirow{2}{*}{line} & ~~$2\cos(k_x+k_z)-\cos(k_x/2-\sqrt{3}k_y/2-k_z)-\cos(k_x/2+\sqrt{3}k_y/2-k_z)$~~ & ~~$ \phi_A(k_x,k_y,-k_z)$~~ \\
		 &                        &                          &                       & ~~$-\cos(k_x/2-\sqrt{3}k_y/2-k_z)+\cos(k_x/2+\sqrt{3}k_y/2-k_z)$              ~~ & ~~$ \phi_A(k_x,k_y,-k_z)$~~ \\
		 & \multirow{2}{*}{$E_2$} & \multirow{2}{*}{triplet} & \multirow{2}{*}{line} & ~~$-\sin(k_x/2-\sqrt{3}k_y/2-k_z)+\sin(k_x/2+\sqrt{3}k_y/2-k_z)$              ~~ & ~~$-\phi_A(k_x,k_y,-k_z)$~~ \\
		 &                        &                          &                       & ~~$2\sin(k_x+k_z)+\sin(k_x/2-\sqrt{3}k_y/2-k_z)+\sin(k_x/2+\sqrt{3}k_y/2-k_z)$~~ & ~~$-\phi_A(k_x,k_y,-k_z)$~~ \\
    \hline
  \end{tabular}
  \label{table:Irrep}
\end{table*}

In the above experiments\cite{Alpern_NJP_2016, Alpern_NL_2019, Alpern_PRM_2021}, the helical molecules have been absorbed on the Nb substrate.
In this paper, we focus on the possibility that the pair potentials are induced in the helical molecules by the superconducting proximity effect.
In this scenario, we have to clarify what types of pairings are induced and how they generate the bound states on the surface of helical lattices.
For this purpose, we adopt the model calculation of helical crystals and investigate the surface bound states on the (001) and zigzag surfaces of helical lattices for all the possible nearest interlayer pairings, where 
the (001) surface perpendicular to the helical axis corresponds to the edge of the helical molecules observed by the STS and STM experiments.  
For $A_1$ and $E_1$ representations of spin-singlet and $E_2$ representation of spin-triplet, zero energy peaks in the surface density of states (SDOS) are obtained on the (001) surface.
For $E_1$ and $E_2$ representations, zero energy peaks are obtained on the zigzag surface.
In addition, we verify that the corresponding winding numbers are non-trivial. 

This paper is organized as follows: 
In Sec. \ref{form}, we introduce the tight-binding model for the helical lattices, the recursive Green function method and one dimensional (1D) winding number.
In Sec. \ref{subsec:Irrep}, we classify the possible pair potentials into the irreducible representations of the point group.
In Sec. \ref{subsec:SDOS}, we show the numerical results of the SDOS.
In Sec. \ref{subsec:1DWN}, we show the numerical results of the winding number and verify the consistency between the appearance of zero energy peaks in the SDOS and non-trivial winding number.
We summarize our results in Sec. \ref{subsec:con}.

\section{FORMULATION}\label{form}

In this paper, we consider a three-dimensional (3D) helical lattice with the $D_{6}$ point groups as shown in Fig.~\ref{fig:lattice}. We also consider the 3D honeycomb lattice with $D_{6h}$ as a reference.
The helical and 3D honeycomb lattice is composed of a stack of honeycomb lattice layers, which have two sub-lattice A and B in each unit cell.  
The unit cell in our model is spanned by the primitive vector $\bm{a}_1=a\hat{\bm{x}}$, $\bm{a}_2=a/2(-\hat{\bm{x}}+\sqrt{3}\hat{\bm{y}})$ and $\bm{a}_3=a\hat{\bm{z}}$ shown in Figs.~\ref{fig:lattice}(b) and (d), 
where $\hat{\bm{x}}$, $\hat{\bm{y}}$ and $\hat{\bm{z}}$ are unit vectors along $x$, $y$ and $z$ axis, and $a$ is a lattice constant. 
Thus, each unit cell is labeled by a vector of integer $\bm{n}=(n_1,n_2,n_3)$, 
where the A (B) site in a unit cell $\bm{n}$ is located at $\sum_i n_i\bm{a}_i$ ($\sum_i n_i\bm{a}_i+a\hat{\bm{y}}/\sqrt{3}$).
On this lattice, we examine the superconducting state with the interlayer pairings in the tight-binding model on the helical and 3D honeycomb lattices. 
We consider both spin-singlet and triplet pairings.
With respect to the triplet pairings, it is sufficient to consider the antiparallel spin pairings because of the spin rotational symmetry in the present system.
The corresponding Hamiltonian $\hat{H}$ is given by\cite{Yoda_SR_2015,Yoda_NL_2018}
\begin{align}
	\hat{H}&=t_1\sum_{\braket{ij}\sigma}\hat{c}^\dagger_{i\sigma}\hat{c}_{j\sigma}+t_2\sum_{[ij]\sigma}\hat{c}^\dagger_{i\sigma}\hat{c}_{j\sigma}+t_3\sum_{\{ij\}\sigma}\hat{c}^\dagger_{i\sigma}\hat{c}_{j\sigma}\nonumber
	     \\&+\sum_{ij}[\Delta_{ij}\hat{c}_{i\uparrow}\hat{c}_{j\downarrow}+\mathrm{h.c.}]\label{eq:helical},
\end{align}
where $c_{i\sigma}$ ($c_{i\sigma}^\dagger$) is an annihilation (creation) operator for an electron with the spin $\sigma$ at the site $i$, 
$t_1$, $t_2$ and $t_3$ are hopping amplitudes, and $\Delta_{ij}$ is the pair potential of the superconductivity. 
The site $i$ indicates the set of the unit cell $\bm{n}$ and the sub-lattice A or B.
In our paper, the chemical potential is set to zero.
The first term in Eq.~\eqref{eq:helical} represents a nearest-neighbor hopping in $xy$ plane. 
The second and third terms represent nearest neighbor layer hoppings in the helical and honeycomb lattices, respectively,
as shown in Figs.~\ref{fig:lattice}(b) and (c). 
We set $t_3$ ($t_2$) to zero when we consider the helical (honeycomb) lattice.
We consider the nearest neighbor layer pairings depending on $k_z$ to investigate the pair potentials generating the bound states on the (001) surface as the blue plane in Fig.~\ref{fig:surface}(a).
We also investigate the bound states on the zigzag surface as the red plane in Fig.~\ref{fig:surface}(b).
In the interlayer pairings, $\Delta_{ij}$ only has a finite value when the set of $i$ and $j$ belongs to the same sub-lattice.
Due to the spin-rotational symmetry, it is sufficient to consider the anti-parallel spin pairings, i.e., the Cooper pairings has zero total spin on the quantization axis.
In the last term of Eq.~\eqref{eq:helical}, the spin-singlet (spin-triplet) states corresponds to the pair potential which satisfies $\Delta_{ij}=\Delta_{ji}$ ($\Delta_{ij}=-\Delta_{ji}$). 

The Brillouin zone and high symmetry points are shown in Fig.~\ref{fig:hsp}.
	The K points $\bm{b}_i$ is defined as $\bm{b}_1\cdot\bm{a}_1=-\bm{b}_2\cdot\bm{a}_2=\bm{b}_3\cdot(\bm{a}_2-\bm{a}_1)=-2\pi/3$ on $k_z=0$ plane, and the $K^\prime$, $H$ and $H^\prime$ points are defined as $-\bm{b}_i$, $\bm{b}_i+\pi\hat{\bm{z}}/a$ and $-\bm{b}_i+\pi\hat{\bm{z}}/a$, respectively.

We calculate the SDOS at the (001) and zigzag surfaces of semi-infinite helical and honeycomb lattices.
For this purpose, we consider the clean system with the (001) and zigzag surfaces as shown in Figs.~\ref{fig:surface} (a) and (b),
where we assume the periodic boundary condition along the direction parallel to the surface.
Thus, the system is described by the momentum parallel to the surface $\bm{k}_\parallel$ and integer $n_\perp$ specifying the layers stacked along $\bm{a}_\perp$ direction, where $\bm{k}_\parallel$, $n_\perp$ and $\bm{a}_\perp$ are given by
$\bm{k}_\parallel=k_x\hat{\bm{x}}+k_y\hat{\bm{y}}$, $n_\perp=n_3$ and $\bm{a}_\perp=\bm{a}_3$
($\bm{k}_\parallel=k^\prime(\hat{\bm{x}}/2+\sqrt{3}\hat{\bm{y}}/2)+k_z\hat{\bm{z}}$, $n_\perp=n_1$ and $\bm{a}_\perp=\bm{a}_1$) in the system with the (001) (zigzag) surface, respectively.
The surface Brillouin zones projected to the (001) and zigzag surfaces are shown in Figs.~\ref{fig:surface} (c) and (d), respectively.
The layers in the SC are labeled from $n_\perp=1$ to $\infty$, and the layer $n_\perp=1$ corresponds to the surface.
This means that the problem is reduced to the one dimensional problem along the direction $\bm{a}_\perp$ at each momentum $\bm{k}_\parallel$.
The Hamiltonian $\hat{H}$ is written as
\begin{align}
	\hat{H}&=\frac{1}{2}\sum_{\bm{k}_\parallel}\sum_{n_\perp, n_\perp^\prime}\hat{\Psi}^\dagger_{n_\perp}(\bm{k}_\parallel)\tilde{H}_{n_\perp n_\perp^\prime}(\bm{k}_\parallel)\hat{\Psi}_{n_\perp^\prime}(\bm{k}_\parallel),\nonumber\\
\tilde{H}&_{n_\perp n_\perp^\prime}(\bm{k}_\parallel)
	=\begin{pmatrix}
		\hat{h}_{n_\perp n_\perp^\prime}(\bm{k}_\parallel)\hat{s}_0 & -\hat{\Delta}^*_{n_\perp n_\perp^\prime}(\bm{k}_\parallel)i\hat{s}_y \\
	\hat{\Delta}_{n_\perp n_\perp^\prime}(\bm{k}_\parallel)i\hat{s}_y & -\hat{h}_{n_\perp n_\perp^\prime}(\bm{k}_\parallel)\hat{s}_0
\end{pmatrix},\nonumber\\
\hat{\Psi}&_{n_\perp}(\bm{k}_\parallel)=(\hat{\bm{C}}_{n_\perp}(\bm{k}_\parallel),~\hat{\bm{C}}^*_{n_\perp}(-\bm{k}_\parallel))
\end{align}
where $\hat{\cdot}$ is a 2 $\times$ 2 matrix in the sub-lattice space, 
$n_\perp$ is a label of the layer parallel to the surface,
$\hat{s}_i$ ($i=0,x,y,z$) is the Pauli matrix acting on the spin space and 
$\hat{\bm{C}}_{n_\perp}(\bm{k}_\parallel)=(    $  
$\hat{c}_{n_\perp\bm{k}_\parallel A\uparrow  },$
$\hat{c}_{n_\perp\bm{k}_\parallel A\downarrow},$
$\hat{c}_{n_\perp\bm{k}_\parallel A\uparrow  },$
$\hat{c}_{n_\perp\bm{k}_\parallel A\downarrow},$
$\hat{c}_{n_\perp\bm{k}_\parallel B\uparrow  },$
$\hat{c}_{n_\perp\bm{k}_\parallel B\downarrow},$
$\hat{c}_{n_\perp\bm{k}_\parallel B\uparrow  },$
$\hat{c}_{n_\perp\bm{k}_\parallel B\downarrow} $
$)$ is a spinor composed of annihilation operators $\hat{c}_{n_\perp\bm{k}_\parallel \mu\sigma} $ of the electrons with spin $\sigma$, momentum $\bm{k}_\parallel$ and sub-lattice $\mu$ at the $n_\perp$th layer.

The Green's function at the $n_\perp$ and $n_\perp^\prime$th layers, spins $\sigma$ and $\sigma^\prime$ and sub-lattice $\mu$ and $\mu^\prime$ and the complex frequency $\omega$ is defined as follows:
\begin{align}
	\tilde{G}(\bm{k}_\parallel,\omega)=\left(\omega\tilde{I}-\tilde{H}(\bm{k}_\parallel)\right)^{-1},
\end{align}
where $\tilde{I}$ is a unit matrix  with the same size as $\tilde{H}(\bm{k}_\parallel)$.
The SDOS is calculated from the retarded Green's function:
\begin{align}
	\rho_\sigma(E)&=-\frac{1}{2\pi N_S}\sum_{\mu=A,B}\int\mathrm{Im}\left[G^{1111}_{\mu\sigma\mu\sigma}(\bm{k}_\parallel,E+i\eta)\right]d\bm{k}_\parallel,
\end{align}
where $G^{\tau\tau^\prime n_\perp n_\perp^\prime}_{\mu\sigma\mu^\prime\sigma^\prime}$ is a matrix element of $\tilde{G}$ at the particle-hole indices $\tau$ and $\tau^\prime$, $n_\perp$ and $n_\perp^\prime$th layers, the sub-lattice $\mu$ and $\mu^\prime$ and the spin $\sigma$ and $\sigma^\prime$,
$E$ and $\eta$ are the energy and smearing factor, respectively, $N_S$ is a number of sites on the surface and $n_\perp=1$ shows the layer of the surface. 
As a result of the zero total spin of the Cooper pairs, the surface density of states is independent of the spin $\sigma$, i.e., $\rho(E)=\rho_\uparrow(E)=\rho_\downarrow(E)$.
To calculate the retarded Green's function at the surface $n_\perp=1$,
we apply the recursive Green function method proposed by Umerski\cite{umerski_PRB_1997,yada_JPSJ_2014,takagi_PRB_2020}.

The dispersionless ABSs generated on the surface of anisotropic SCs are characterized by the non-trivial 1D winding number defined in the bulk\cite{sato_PRB_2011}.
The BdG Hamiltonian in the bulk is written as
\begin{align}
	\mathcal{H}(\bm{k})&=\frac{1}{2}
	\begin{pmatrix}
		\check{\varepsilon}(\bm{k}) & \check{\Delta} \\
		\check{\Delta}^\dagger & -\check{\varepsilon}^{T}(-\bm{k})
	\end{pmatrix},\label{BdGhamiltonian}
\end{align}
where $\check{\cdot}$ is a 4 $\times$ 4 matrix in the direct product of the sub-lattice and spin spaces.
Having the time-reversal symmetry, the BdG Hamiltonian satisfies
\begin{align}
	\Theta\mathcal{H}(\bm{k})&\Theta^{-1}=\mathcal{H}^{*}(-\bm{k}),\quad\Theta=\begin{pmatrix}
	i\hat{s}_y\hat{\tau}_0 & 0 \\
	0 & i\hat{s}_y\hat{\tau}_0
  \end{pmatrix},
\end{align}
where $\hat{\tau}_i$ ($i=0,x,y,z$) is the Pauli matrix in the sub-lattice space.
In addition, the BdG Hamiltonian has the particle-hole symmetry written as:
\begin{align}
	\mathcal{C}\mathcal{H}(\bm{k})&\mathcal{C}^{-1}=-\mathcal{H}^{*}(-\bm{k}),\quad\mathcal{C}=
  \begin{pmatrix}
    0 & \hat{s}_0\hat{\tau}_0 \\
    \hat{s}_0\hat{\tau}_0 & 0
  \end{pmatrix}
.
\end{align}
In order to define the winding number, we introduce the chiral operator as $\Gamma=-i\mathcal{C}\Theta$ in the spin-singlet case and $\Gamma=S_z\mathcal{C}\Theta$ in the spin-triplet case\cite{sato_PRB_2011,Kobayashi_PRB_2015}, where $S_z$ is the $z$-component of the spin operator defined as:
\begin{align}
  S_z=\begin{pmatrix}
  	  \hat{s}_{z}\hat{\tau}_0 & 0 \\
	    0 & -\hat{s}_{z}\hat{\tau}_0
  \end{pmatrix}
.
\end{align}
Thus, the flat bands for the triplet pairs are unstable against the spin-orbit interactions.

The 1D winding number manifesting the dispersionless ABSs is defined with $\Gamma$ for $\bm{k}_\parallel$ as:
\begin{align}
	w(\bm{k}_\parallel)&=-\frac{1}{4\pi i}\int d\bm{k}_\perp\mathrm{tr}[\Gamma\mathcal{H}^{-1}(\bm{k})\partial_{\bm{k}_\perp}\mathcal{H}(\bm{k})]\label{winding_number}
\end{align}
where $\bm{k}_\perp$ is a momentum perpendicular to the surface and 
the integration is taken over the possible $\bm{k}_\perp$ on the Brillouin zone. 
The winding number at $\bm{k}_\parallel$ is equal to the integer value $N_+-N_-$, 
where $N_\pm$ is the number of zero energy states with an eigenvalue $\Gamma=\pm1$ at $\bm{k}_\parallel$.


\section{Results}

\begin{figure}[htb]                                        
\parbox{0.48\linewidth}{\centering (a)~$A_2$
\includegraphics[width=\linewidth]{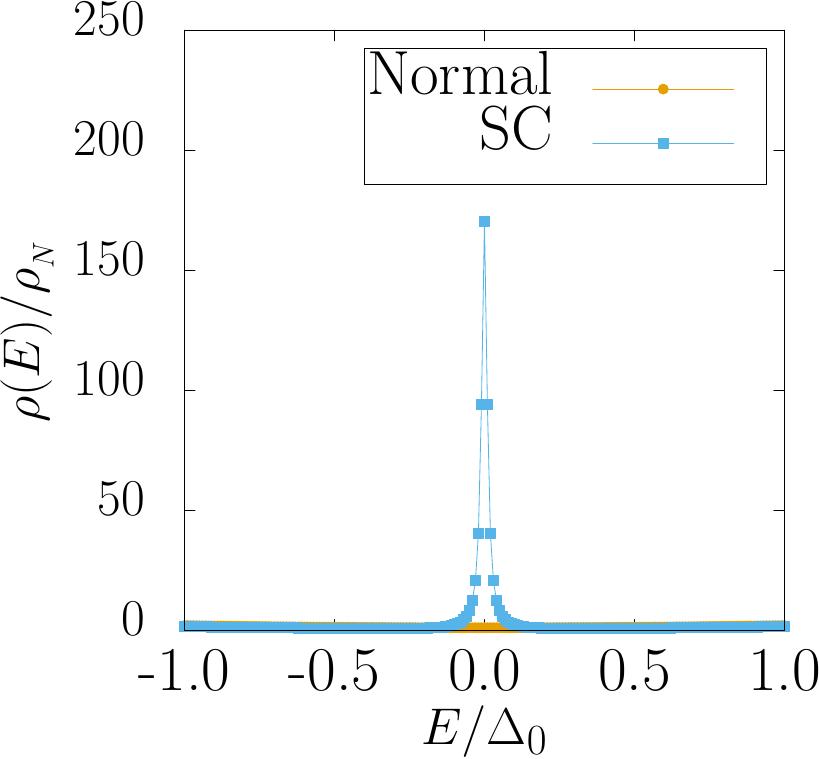}
}
\parbox{0.48\linewidth}{\centering (b)~$A_{2u}$
\includegraphics[width=\linewidth]{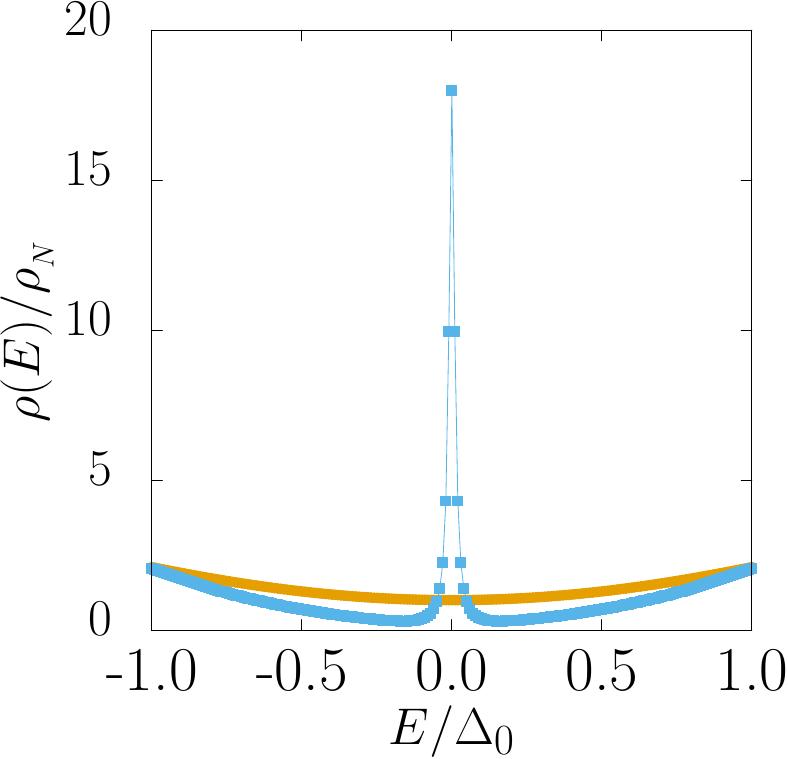}
}
\parbox{0.48\linewidth}{\centering (c)~$E_1$(singlet)
\includegraphics[width=\linewidth]{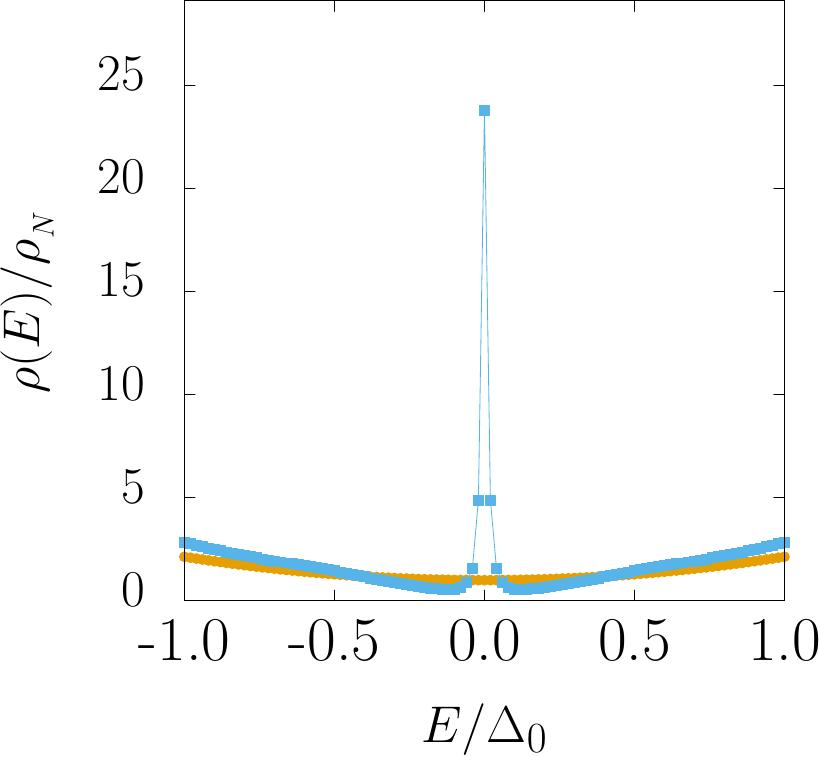}
}
\parbox{0.48\linewidth}{\centering (d)~$E_2$(triplet)
\includegraphics[width=\linewidth]{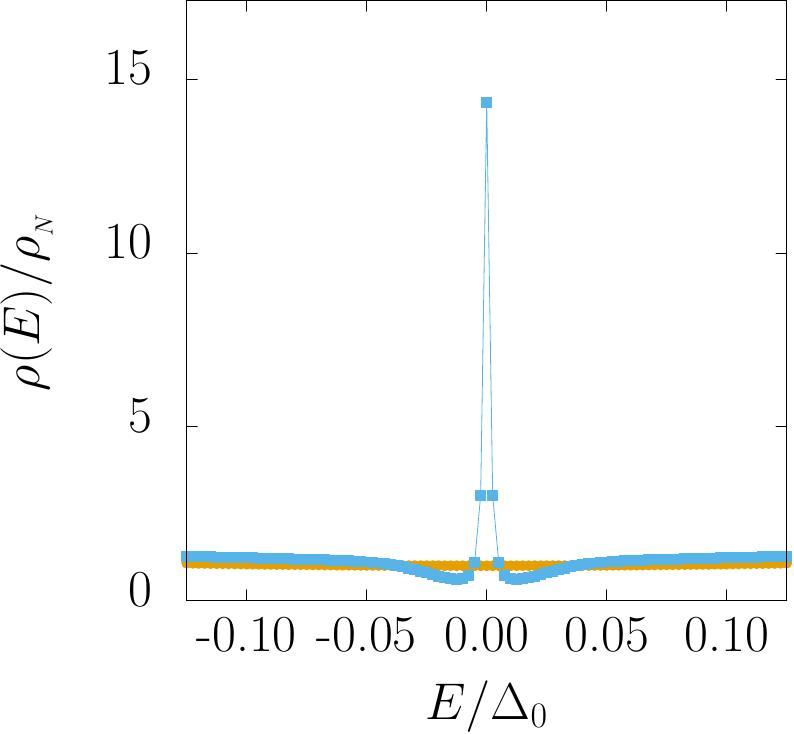}
}
\caption{
	Surface density of states at the (001) surface of the helical and honeycomb lattices in the normal and superconducting states.
	The SDOS are normalized by $\rho_{_N}$ being the zero energy SDOS of the normal state.
	The irreducible representations are shown on top of each figure. 
	The SDOS of $A_{2u}$ is calculated in the honeycomb lattice, and the others are in the helical lattice.
	We specify either spin singlet or triplet for $E_1$ and $E_2$ representations.
	In the irreducible representations that are not shown here, no zero energy peaks appear at the (001) surface. 
	We take $t_1$ as an energy unit and set other hopping integrals as $t_2=0.1$ or $t_3=0.1$. 
	The amplitudes of pair potentials are set as $\Delta_0=0.18$ for $A_2$ and $A_{2u}$, $\Delta_0=0.2$ for $E_1$(singlet) and $\Delta_0=0.4$ for $E_2$(triplet).
	}
\label{fig:SDOS_001}
\end{figure}

\begin{figure}[htb]                                        
\parbox{0.48\linewidth}{\centering (a)$E_1$(singlet)                      
\includegraphics[width=\linewidth]{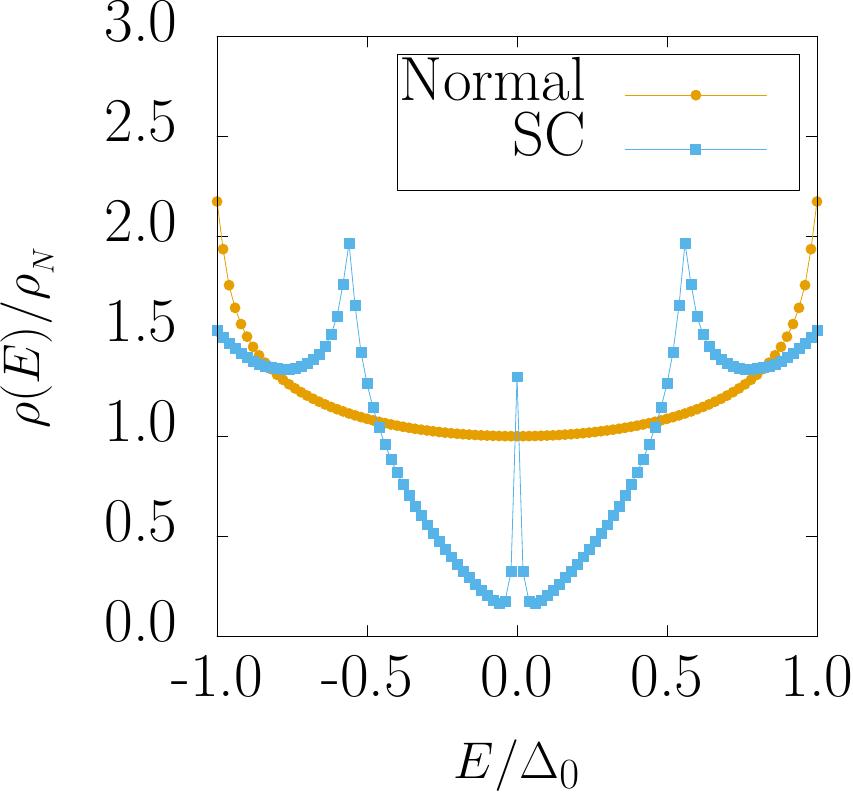}
}
\parbox{0.48\linewidth}{\centering (b)$E_1$(triplet)                  
\includegraphics[width=\linewidth]{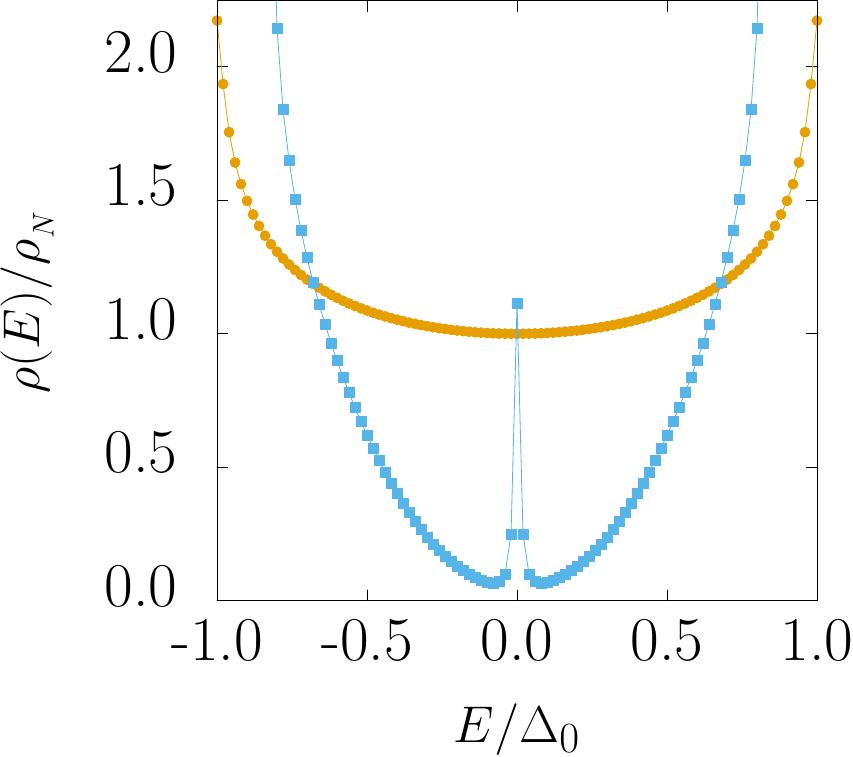}
}
\parbox{0.48\linewidth}{\centering (c)$E_2$(singlet)                 
\includegraphics[width=\linewidth]{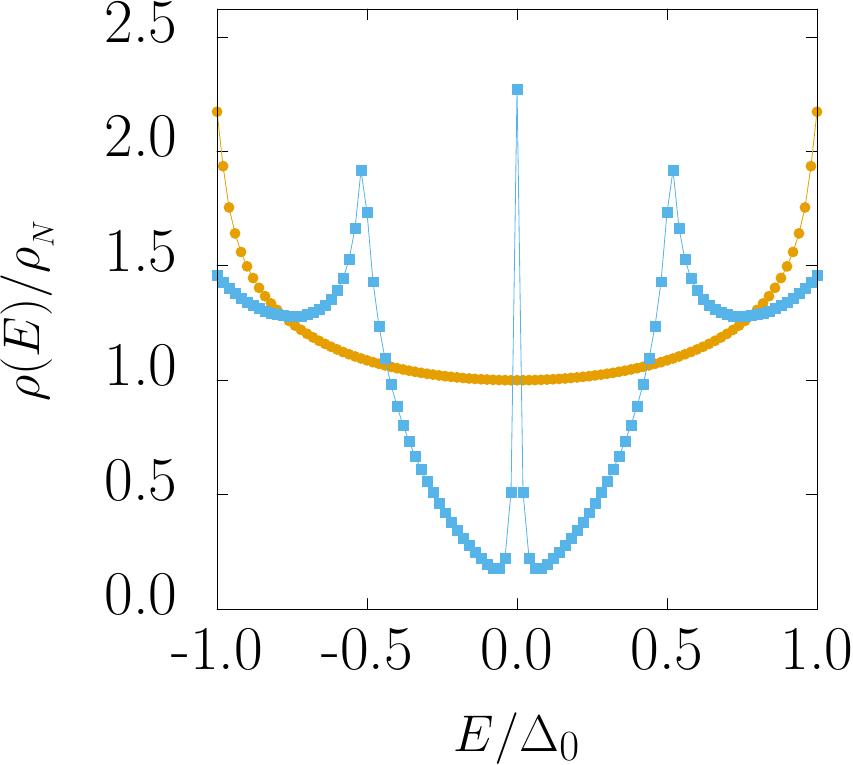}
}
\parbox{0.48\linewidth}{\centering (d)$E_2$(triplet)                 
\includegraphics[width=\linewidth]{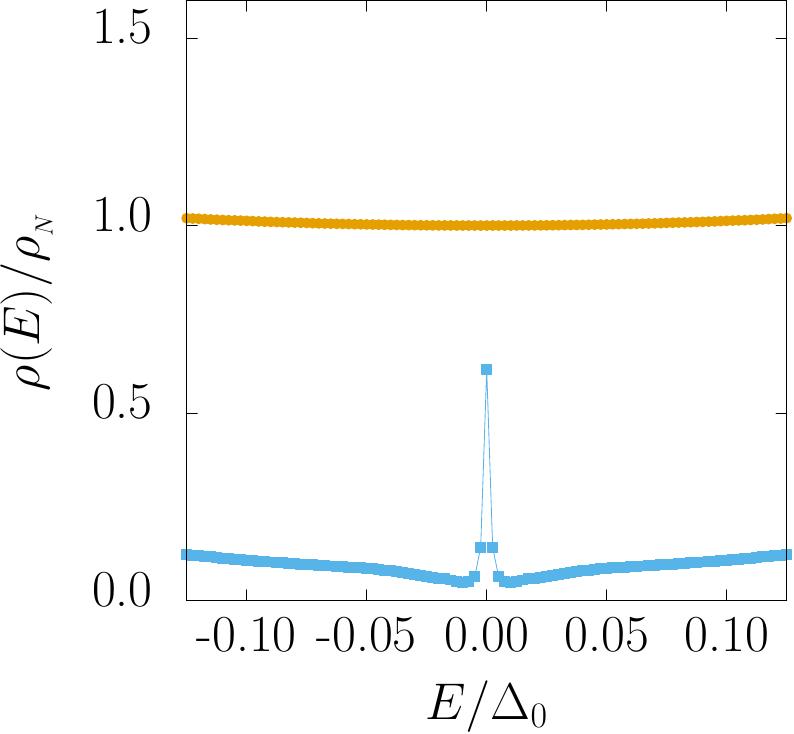}
}
\caption{
	Surface density of states at the zigzag surface of the helical lattice in the normal and superconducting states.
	The figures are shown in the same manner as Fig.~\ref{fig:SDOS_001}.
	In the irreducible representations that are not shown here, no zero energy peaks appear at the zigzag surface. 
	The hopping parameters are set as the same values as in Fig.~\ref{fig:SDOS_001}. 
	The amplitudes of the pair potential are set as $\Delta_0=0.4$ for $E_2$(triplet) and $\Delta_0=0.2$ for the other irreducible representations.
	}
\label{fig:SDOS_zigzag}
\end{figure}

\subsection{Irreducible representations}\label{subsec:Irrep}

We will investigate the possible pair potentials generating the bound states and the resulting surface bound states in the helical lattice.
For this purpose, we consider the nearest layer pairings with the $k_z$ dependence.
In this case, the two electrons on the same sub-lattice constitute the Cooper pair.
Thus, $\Delta_{ij}$ only has a finite value when $i$ and $j$ belong to the same sub-lattice. 

The possible order parameters are classified by the irreducible representations of the point group symmetry\cite{sigrist_RMP_1991}. 
We decompose the pair potentials into the irreducible representations and rewrite the superconducting parts of the Hamiltonian as:
\begin{align}
	\hat{H}_{\Delta}=\Delta_0&\sum_{\bm{k},\mu=A,B}[\phi_\mu^{IR}(\bm{k})\hat{c}_{\mu\bm{k}\uparrow}\hat{c}_{\mu-\bm{k}\downarrow}+h.c.],\label{Eq:irrep}
\end{align}
where $\Delta_0$ is the amplitude of the pair potential, $\bm{k}$ and $\mu$ are the momentum and index of sub-lattice, respectively, and $\phi_{\mu}^{IR}(\bm{k})$ is the basis function of the irreducible representation of $D_6$ or $D_{6h}$. 
The basis functions $\phi_{\mu}^{IR}(\bm{k})$ in the helical lattice with $D_{6}$ and honeycomb lattice with $D_{6h}$ are shown in Table.~\ref{table:Irrep}.
There are two kinds of basis functions distinguished by spin channels in $E_1$ and $E_2$ representations.
Hereafter, when necessary in $E_1$ and $E_2$ representations, we append the spin channel to specify the basis function; for example, we write $E_1$ representation of the spin singlet as $E_1$(singlet). 
There are two basis functions in each $E_1$ and $E_2$ representation as seen in Table.~\ref{table:Irrep}. 
We will use the upper one in the model calculation. We have checked that similar results are obtained for the lower basis function.

\subsection{Surface density of states}\label{subsec:SDOS}
In this subsection, we show the numerical results of the SDOS.
We calculate the SDOS at the (001) and zigzag surfaces for all the possible irreducible representations shown in Table.~\ref{table:Irrep}. 
We choose $t_1$ as a unit of the energy and set interlayer hoppings as $t_2/t_1=0.1$ or $t_3/t_1=0.1$.
In Figs.~\ref{fig:SDOS_001} and \ref{fig:SDOS_zigzag}, we show the SDOS for the irreducible representations exhibiting the zero energy peaks in the SDOS.
The SDOS of the irreducible representations belonging to $D_{6}$ ($D_{6h}$) point group are calculated at the surface of the helical (honeycomb) lattice.
The gap size of $E_2$(triplet) is accidentally much smaller than $\Delta_0$ in our hopping parameters.
Thus, we take $\Delta_0$ of $E_2$(triplet) larger than the ones for the other irreducible representations in Figs.~\ref{fig:SDOS_001} and \ref{fig:SDOS_zigzag}.

The zero energy peaks appear at the (001) surface for $A_2$, $A_{2u}$, $E_1$(singlet), and $E_2$(triplet) representations and zigzag surface for $E_1$ and $E_2$ representations.
For the other irreducible representations not shown in Figs.~\ref{fig:SDOS_001} and \ref{fig:SDOS_zigzag}, zero energy peaks are not obtained in the SDOS (see Appendix~\ref{sec:app}).
In the helical lattice, there are three representations, $A_2$, $E_1$(singlet) and $E_2$(triplet) representations, exhibiting the zero energy peak at the (001) surface. 
On the other hand, $A_{2u}$ representation is the only irreducible representation which shows zero energy peak in the honeycomb lattice.
At the zigzag surface, all of the zero energy peaks in Fig.~\ref{fig:SDOS_zigzag} are obtained in the helical lattice. 
These appearance of the zero energy peaks are characterized by 1D winding number in Eq.~\eqref{winding_number} as discussed in the next subsection.


\begin{figure}[htb]                  
\parbox{0.48\linewidth}{\centering (a)$A_2$    
\includegraphics[width=\linewidth]{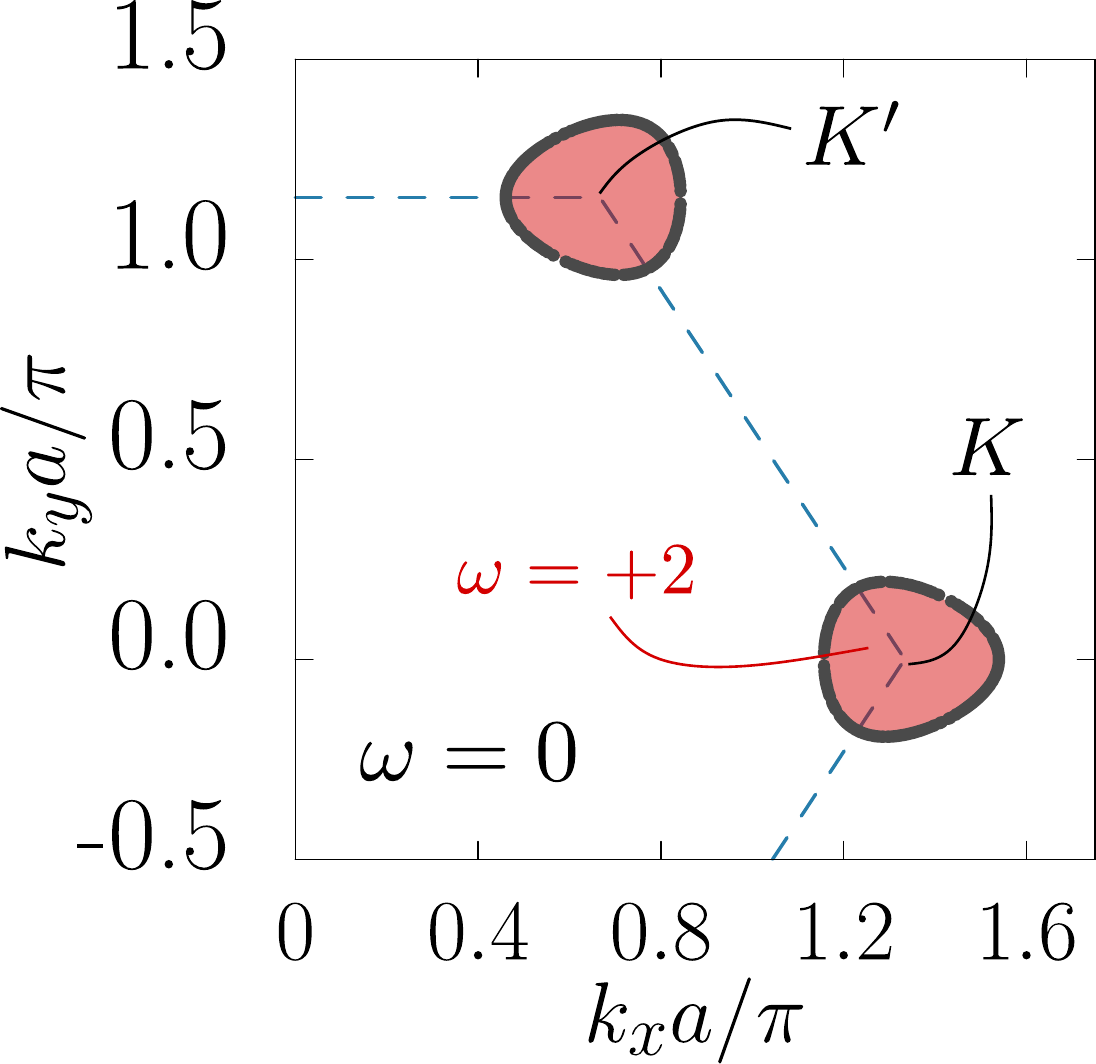}
}
\parbox{0.48\linewidth}{\centering (b)$A_{2u}$        
\includegraphics[width=\linewidth]{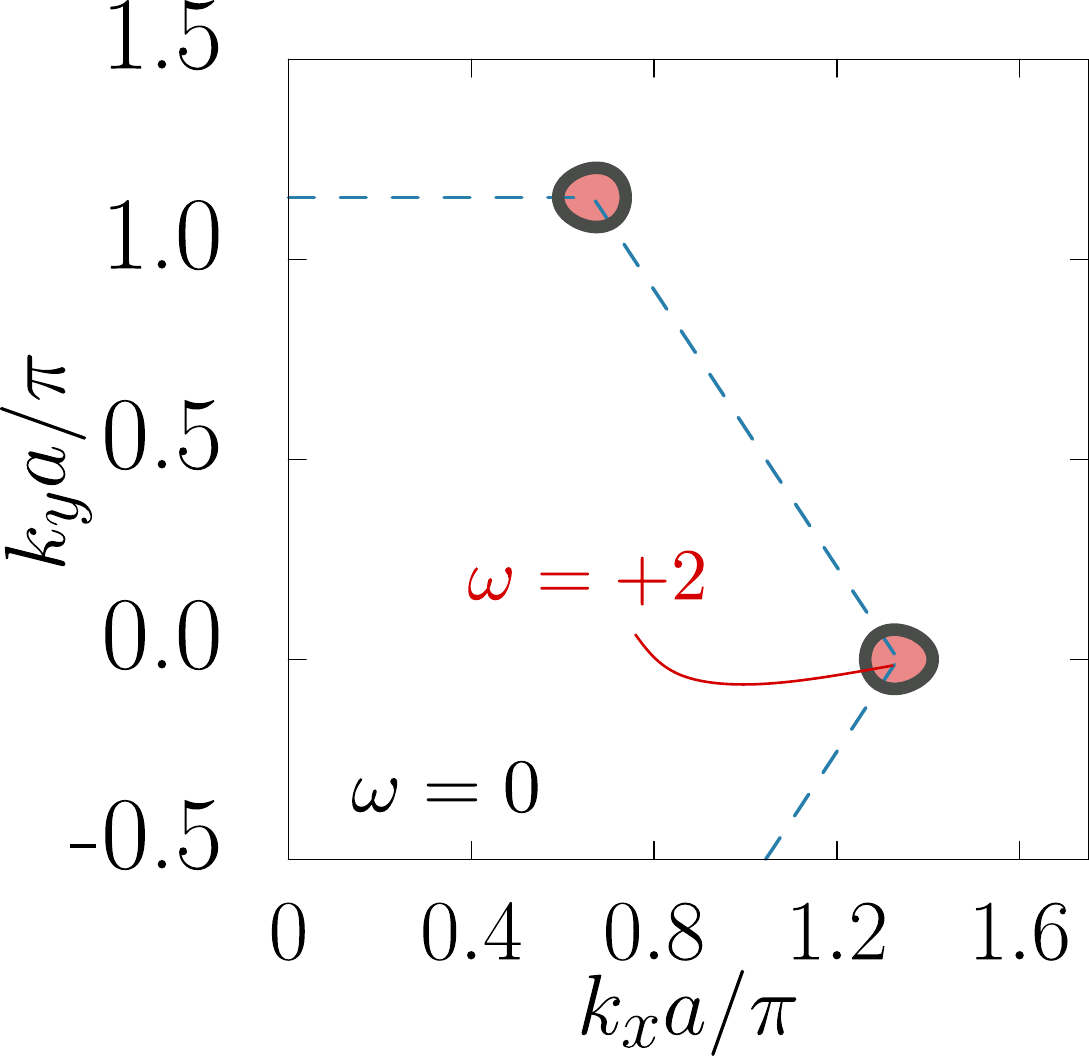}
}
\parbox{0.48\linewidth}{\centering (c)$E_1$(singlet)                           
\includegraphics[width=\linewidth]{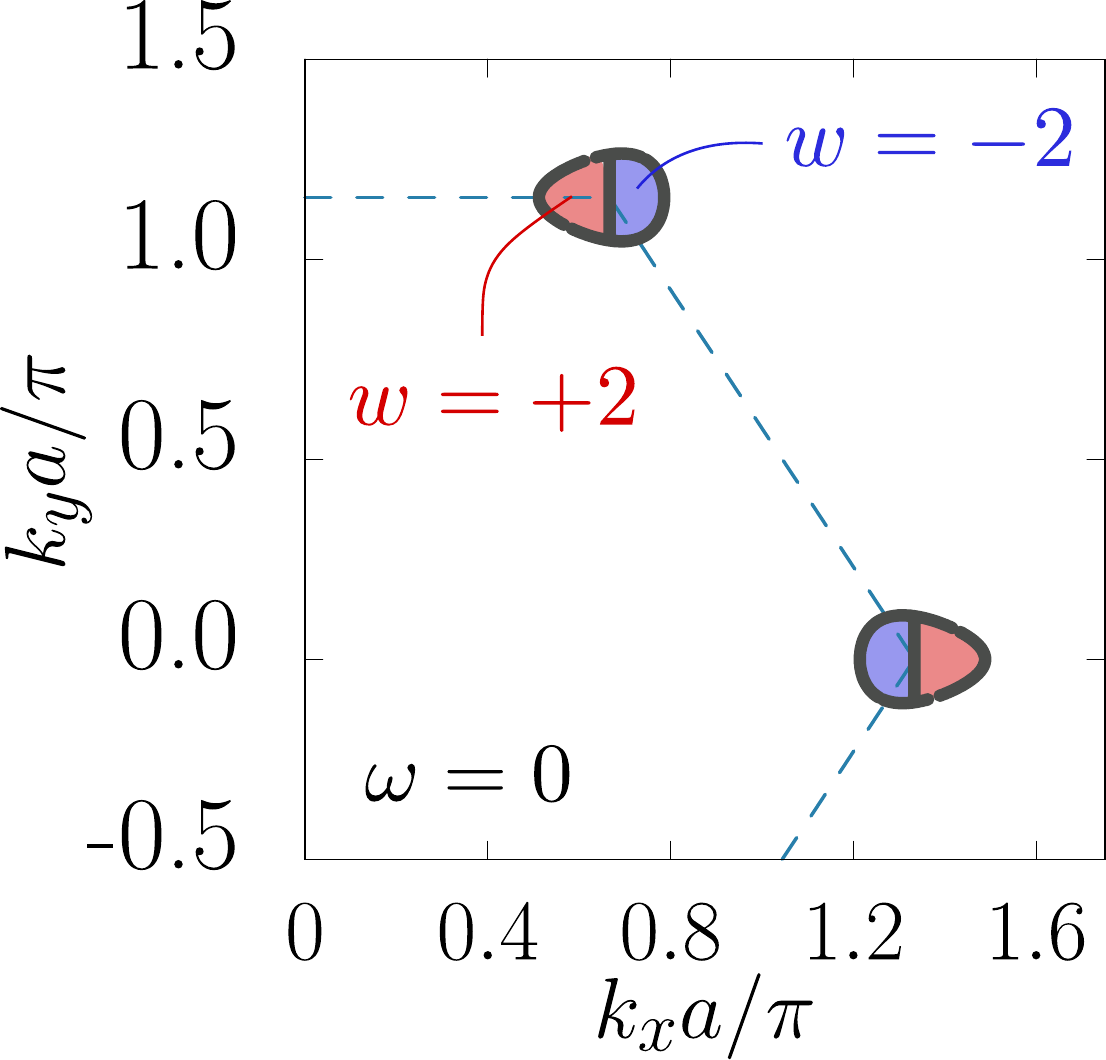}
}
\parbox{0.48\linewidth}{\centering (d)$E_2$(triplet)
\includegraphics[width=\linewidth]{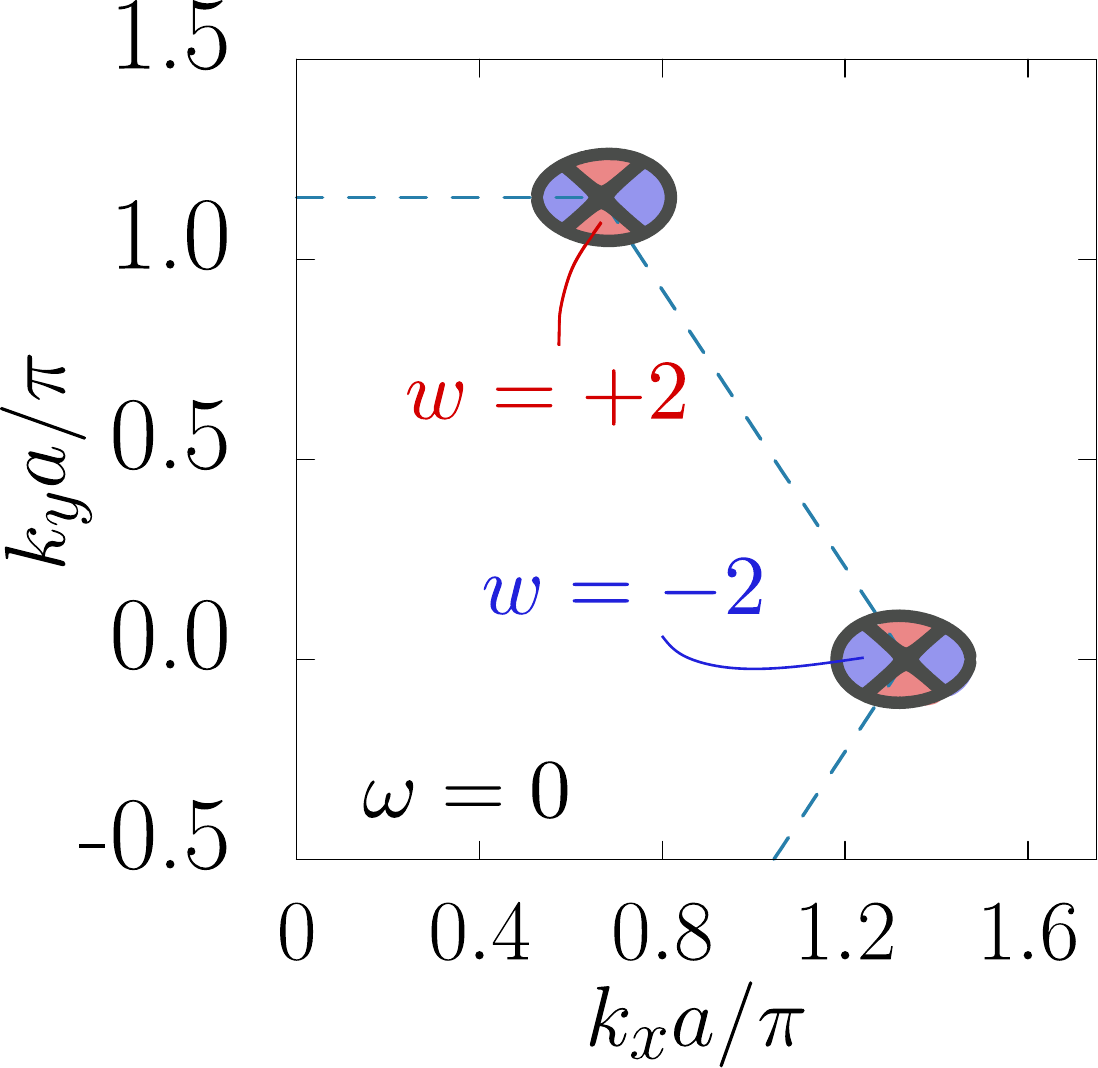}
}
\caption{One dimensional winding number, Eq.~\eqref{winding_number}, as a function of momentum $\bm{k}_\parallel$ parallel to the (001) surface.
	The irreducible representations are shown on top of each figure.
	We show the winding number for the irreducible representations shown in Fig.~\ref{fig:SDOS_001}, and those for the other irreducible representations are zero over the surface momentum $\bm{k}_\parallel$.
	The dashed lines show the boundary of the Brillouin zone projected to the (001) surface.
	For each irreducible representation, we set the hopping parameters and pair potential as the same values as in Fig.\ref{fig:SDOS_001}.
	Red and blue regions indicate $w=\pm2$, while white region represents $w=0$.
	The black lines are the nodes projected on the (001) surface.
	The nodal lines are drawn by plotting the momenta at which $\sqrt{\det[\mathcal{H}_{4\times4}(\bm{k})]}$ is less than $10^{-7}t_1^2$, where $\mathcal{H}_{4\times4}(\bm{k})$ is BdG Hamiltonian reduced to the $4\times4$ matrix.
}
\label{fig:1DWN_001}
\end{figure}

\begin{figure}[htb]                                        
\parbox{0.48\linewidth}{\centering (a)$E_1$(singlet)                      
\includegraphics[width=\linewidth]{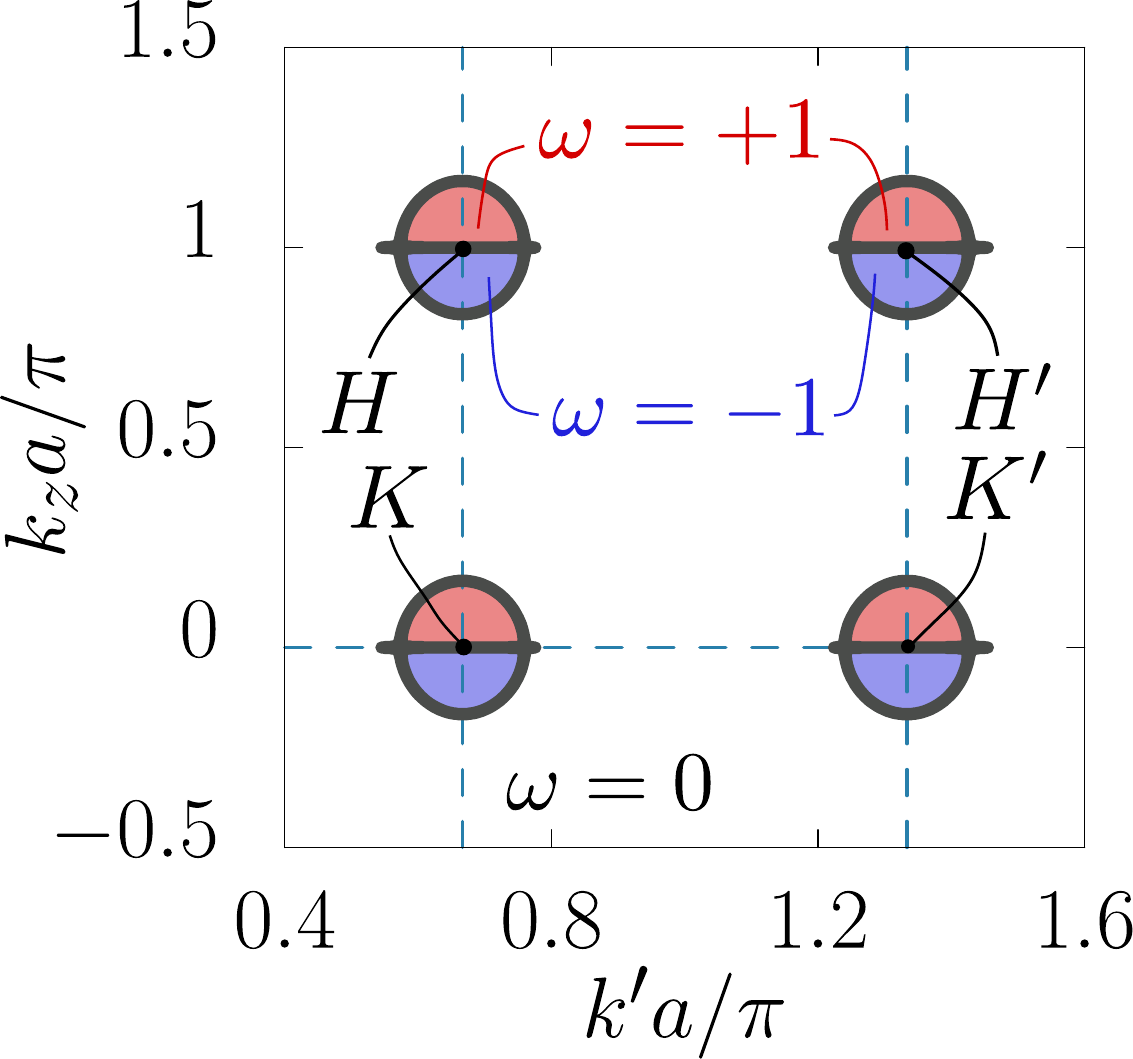}
}
\parbox{0.48\linewidth}{\centering (b)$E_1$(triplet)    
\includegraphics[width=\linewidth]{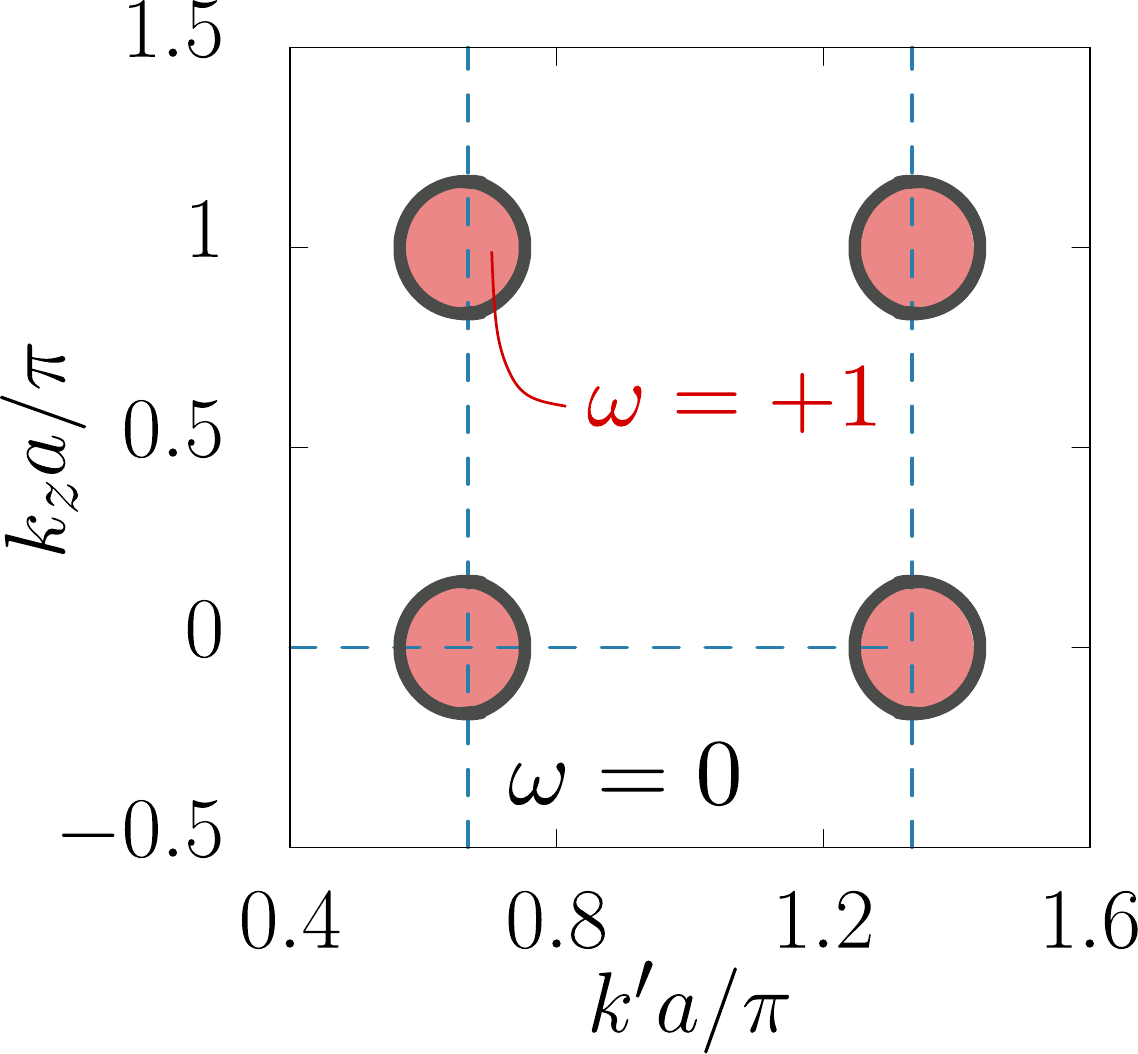}
}
\parbox{0.48\linewidth}{\centering (c)$E_2$(singlet)
\includegraphics[width=\linewidth]{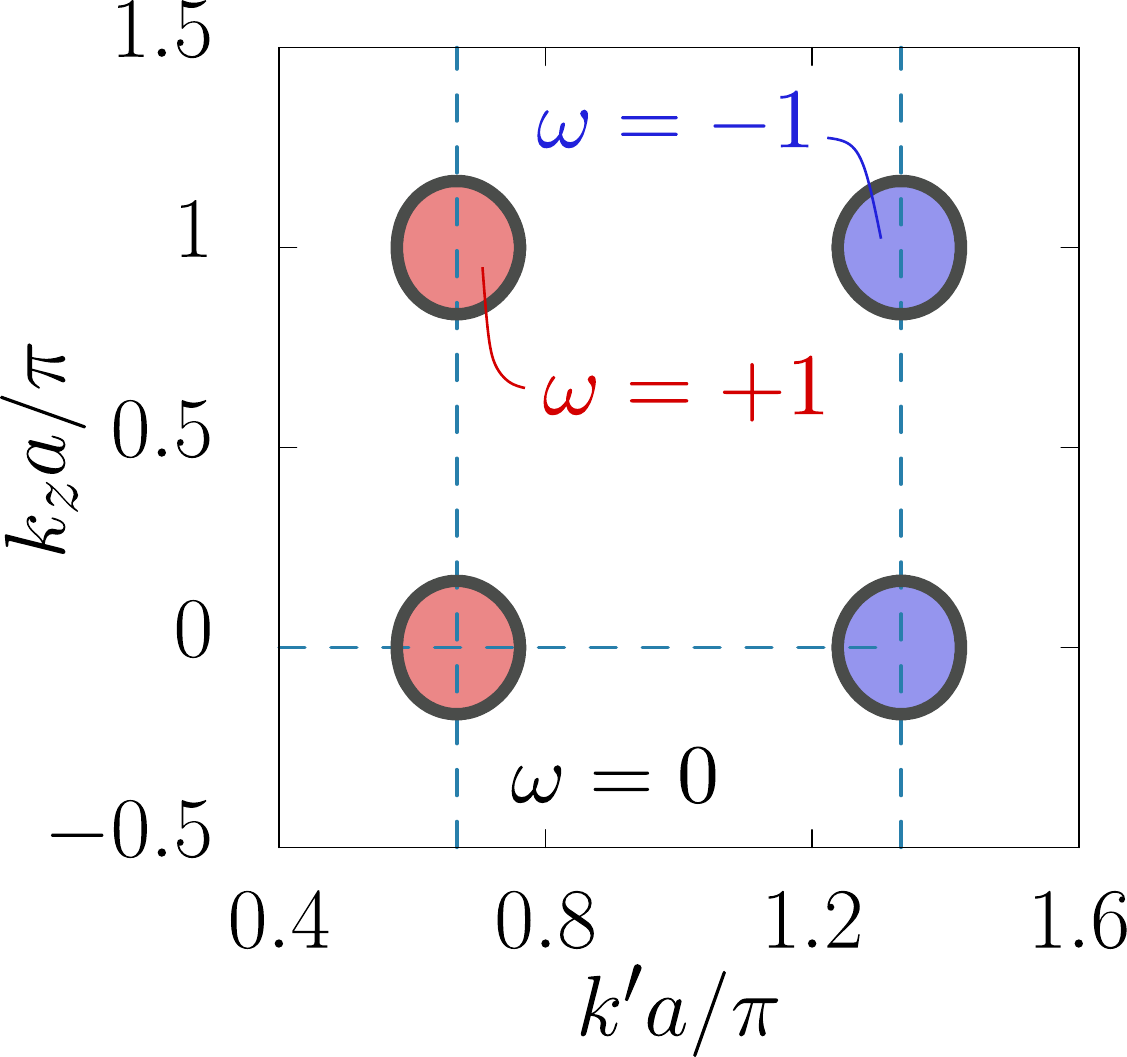}
}
\parbox{0.48\linewidth}{\centering (d)$E_2$(triplet)
\includegraphics[width=\linewidth]{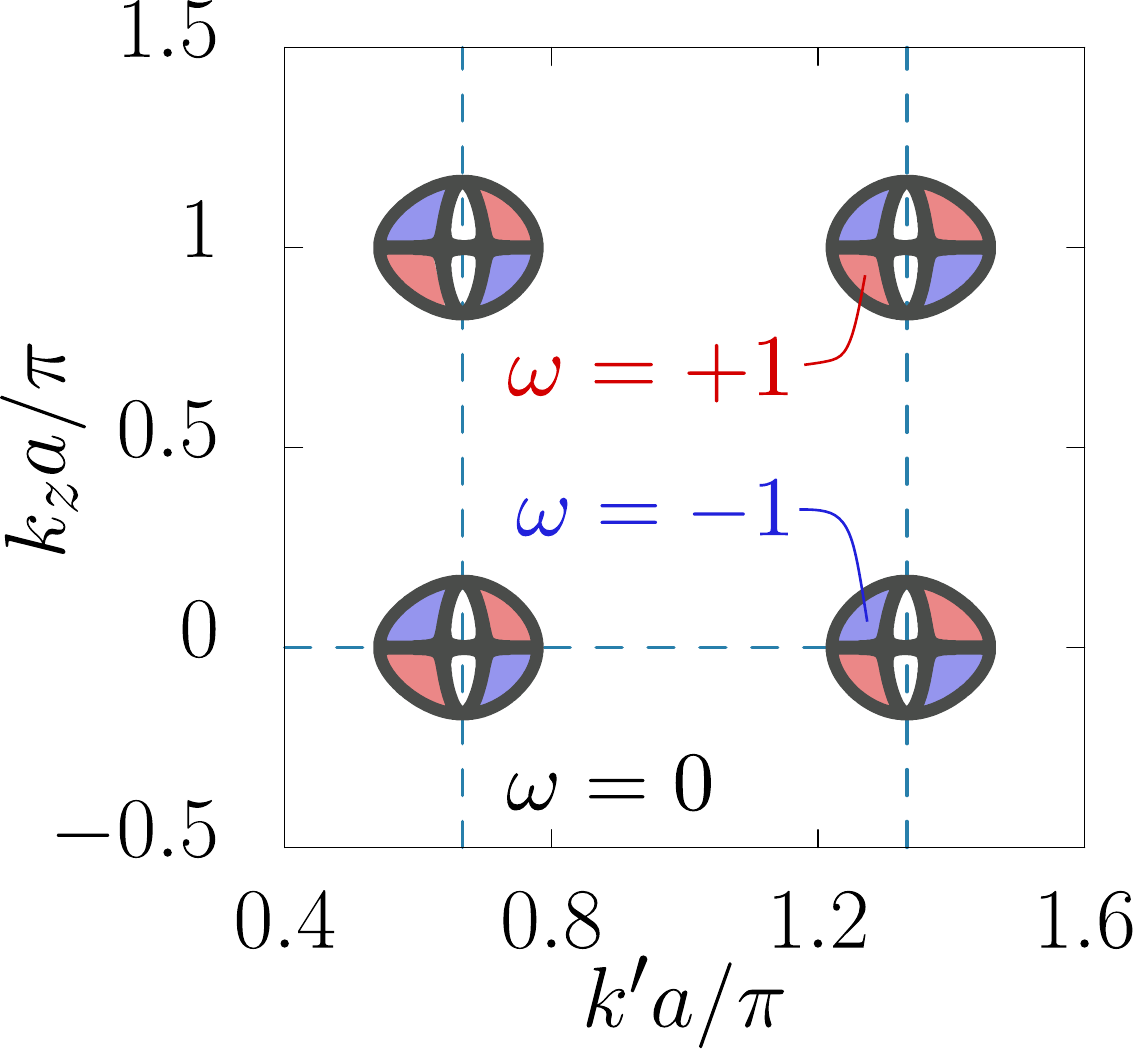}
}
\caption{One dimensional winding number, Eq.~\eqref{winding_number}, as a function of momentum $\bm{k}_\parallel$ parallel to the zigzag surface.
	The irreducible representations are shown on top of each figure.
	We show the winding number for the irreducible representations shown in Fig.~\ref{fig:SDOS_zigzag}, and those for the other irreducible representations are zero over the surface momentum $\bm{k}_\parallel$.
	The dashed lines connect the high symmetry points projected on the zigzag surface.
	For each irreducible representation, we set the hopping parameters and pair potential as the same values as in Fig.\ref{fig:SDOS_zigzag}.
	Red and blue regions indicate $w=\pm1$, while white represents $w=0$.
	The black lines are the line nodes projected on the zigzag surface.
	The nodal lines are drawn by plotting the momenta at which $\sqrt{\det[\mathcal{H}_{4\times4}(\bm{k})]}$ is less than $10^{-7}t_1^2$, where $\mathcal{H}_{4\times4}(\bm{k})$ is BdG Hamiltonian reduced to the $4\times4$ matrix.
}
\label{fig:1DWN_zigzag}
\end{figure}

\subsection{One dimensional winding number}\label{subsec:1DWN}

In this subsection, we calculate the 1D winding number and investigate the correspondence between the presence of the zero energy peaks and flat-band ABSs.
In the numerical calculation of the winding number, because of the spin-rotational symmetry, we reduce the $8\times8$ BdG Hamiltonian $\mathcal{H}(\bm{k})$ in Eq.~\eqref{BdGhamiltonian} to a $4\times4$ matrix $\mathcal{H}_{4\times4}(\bm{k})$.
Thus, the winding numbers shown in this subsection take half of the values defined in Eq.~\eqref{winding_number}.
For all the irreducible representations, we calculate the 1D winding number in the Brillouin zone projected on the (001) and zigzag surfaces.
The hopping parameters are chosen as $t_2/t_1=0.1$ or $t_3/t_1=0.1$ for all the irreducible representations.
For these parameters, the Fermi surfaces are located around high symmetry points $K$, $K^\prime$, $H$ and $H^\prime$ shown in Fig.\ref{fig:lattice}(d).
The winding numbers for the irreducible representations considered in Figs.~\ref{fig:SDOS_001} and \ref{fig:SDOS_zigzag} are shown in Figs.~\ref{fig:1DWN_001} and \ref{fig:1DWN_zigzag}, respectively.
The winding number for the irreducible representations not shown in Figs.~\ref{fig:1DWN_001} and \ref{fig:1DWN_zigzag} is zero over the surface Brillouin zone.

As shown in Fig.~\ref{fig:1DWN_001}, the nodes of the gap function for $A_2$, $A_{2u}$, $E_1$(singlet) and $E_2$(triplet) representations make closed loops around $K$ and $K^\prime$ points in the Brillouin zone projected on the (001) surface. 
In addition, a single nodal line goes through the $K$ and $K^\prime$ points for $E_1$(singlet) representation as shown in Fig.\ref{fig:1DWN_001}(c), and two nodal lines go through these points for $E_2$(triplet) representation as shown in Fig.\ref{fig:1DWN_001}(d).
The winding number has the same value in a region surrounded by the nodal lines and can change across the nodal line. 
The winding number takes $w=+2$ for $A_2$ and $A_{2u}$ representations, and $w=\pm2$ for $E_1$(singlet) and $E_2$(triplet) representations. 

For the irreducible representations shown in Fig.~\ref{fig:1DWN_zigzag}, the nodal lines surround the $K$, $K^\prime$, $H$ and $H^\prime$ points projected on the zigzag surface.
In particular, as shown in Figs.\ref{fig:1DWN_zigzag}(a) and (d), there are two and three nodal lines around these high symmetry points in $E_1$(singlet) and $E_2$(triplet) representations, respectively.
The winding number takes $w=+1$ in Fig.~\ref{fig:1DWN_zigzag}(b) and $w=\pm1$ in the other panels of Fig.~\ref{fig:1DWN_zigzag}.

The non-trivial values of the winding number obtained in this subsections is consistent with the appearance of the zero energy peaks in the SDOS shown in Figs.~\ref{fig:SDOS_001} and \ref{fig:SDOS_zigzag}.
Thus, the zero energy peaks  shown in Figs.~\ref{fig:SDOS_001} and \ref{fig:SDOS_zigzag} originate from the flat band ABSs protected by the topological number.\cite{sato_PRB_2011}
There are three irreducible representations, ${A_1}$, ${E_1}$(singlet) and $E_2$ (triplet) representations, generating the ABSs at the (001) surface of the helical lattice,
and four irreducible representations, ${E_1}$ and $E_2$ representations, generating the ABSs at the zigzag surface of the helical lattice.


\section{CONCLUSION}\label{subsec:con}

\begin{table}[hbtp]
\label{table:conclusion}
\centering
\caption{
	Summary of the results in the helical lattice with $D_{6}$ and honeycomb lattice with $D_{6h}$. 
	The basis functions of each irreducible representation (Irrep) of each point group (PG) are shown in Table.~\ref{table:Irrep}.
	We clarify the spin channels of $E_1$ and $E_2$ to distinguish the basis functions.  
	Checks and crosses indicate the presence and absence of the zero energy peak, respectively. 
	The zero and finite numbers show the trivial and non-trivial winding number.
}
  \begin{tabular}{cccccc}
    \hline 
    PG  & Irrep & \multicolumn{2}{c}{Zero energy peak} & \multicolumn{2}{c}{Winding number} \\
    \hline       &        &   Zigzag  &   (001)  &   Zigzag  &   (001)  \\
		 &        & surface  & surface & surface  & surface \\
    \hline\hline$\mathrm{D_{6h}}$ & $\mathrm{A_{1g}}$               & $\times$     & $\times$     &    $0$ &  $0$   \\
                $               $ & $\mathrm{A_{2u}}$               & $\times$     & $\checkmark$ &    $0$ &  $2$  \\
                $               $ & $\mathrm{B_{2g}}$               & $\times$     & $\times$     &    $0$ &  $0$   \\
                $               $ & $\mathrm{B_{1u}}$               & $\times$     & $\times$     &    $0$ &  $0$   \\
                $\mathrm{D_{6 }}$ & $\mathrm{A_1   }$               & $\times$     & $\times$     &    $0$ &  $0$   \\
                $               $ & $\mathrm{A_2   }$               & $\times$     & $\checkmark$ &    $0$ &  $2$  \\
                $               $ & $\mathrm{B_1   }$               & $\times$     & $\times$     &    $0$ &  $0$   \\
                $               $ & $\mathrm{B_2   }$               & $\times$     & $\times$     &    $0$ &  $0$   \\
                $               $ & \multirow{2}{*}{$E_1$(singlet)} & $\checkmark$ & $\checkmark$ & $\pm1$ & $\pm2$ \\
                $               $ &                                 & $\checkmark$ & $\checkmark$ & $\pm1$ & $\pm2$ \\
                $               $ & \multirow{2}{*}{$E_1$(triplet)} & $\checkmark$ & $\times$     &   $+1$ &  $0$   \\
                $               $ &                                 & $\checkmark$ & $\times$     &   $-1$ &  $0$   \\
                $               $ & \multirow{2}{*}{$E_2$(singlet)} & $\checkmark$ & $\times$     & $\pm1$ &  $0$   \\
                $               $ &                                 & $\checkmark$ & $\times$     & $\pm1$ &  $0$   \\
                $               $ & \multirow{2}{*}{$E_2$(triplet)} & $\checkmark$ & $\checkmark$ & $\pm1$ & $\pm2$ \\
                $               $ &                                 & $\checkmark$ & $\checkmark$ & $\pm1$ & $\pm2$ \\
    \hline
  \end{tabular}
  \label{summary}
\end{table}

We have studied superconductivity in the helical lattice with helical interlayer hopping and the 3D honeycomb lattice as a reference.
We have supposed the nearest interlayer pairings under the mean field theory and decomposed the pair potentials into all the irreducible representations.

We have calculated the SDOS at the (001) and zigzag surfaces for all the possible irreducible representations.
At the (001) surface of the helical lattice, the zero energy peaks have appeared in the SDOS for $A_2$, $E_1$(singlet) and $E_2$(triplet) representations.
At the zigzag surface, the zero energy peaks have been obtained for $E_1$ and $E_2$ representations.
Calculating the 1D winding number, we have clarified the ABSs manifested as zero energy peaks.
We have summarized the appearances of zero energy peaks and values of winding number of all the possible irreducible representations in Table.~\ref{summary}.
~~
\\~\\
{\bf ACKNOWLEDGMENTS}

S.Y. would like to take this opportunity to thank the “Nagoya University
Interdisciplinary Frontier Fellowship” supported by Nagoya University and JST, the
establishment of university fellowships towards the creation of science technology innovation,
Grant Number JPMJFS2120.
T.Y. was supported by JSPS KAKENHI Grant Number JP30578216 and the JSPS-EPSRC Core-to-Core program "Oxide Superspin".
Y.T. was supported by Scientific Research (A) (KAKENHI Grant No. JP20H00131) and Scientific Research (B) (KAKENHI No. JP20H01857).
\\~\\

\appendix

\begin{figure}[htb]                                        
\parbox{0.48\linewidth}{\centering (a)~$A_{1g}$
\includegraphics[width=\linewidth]{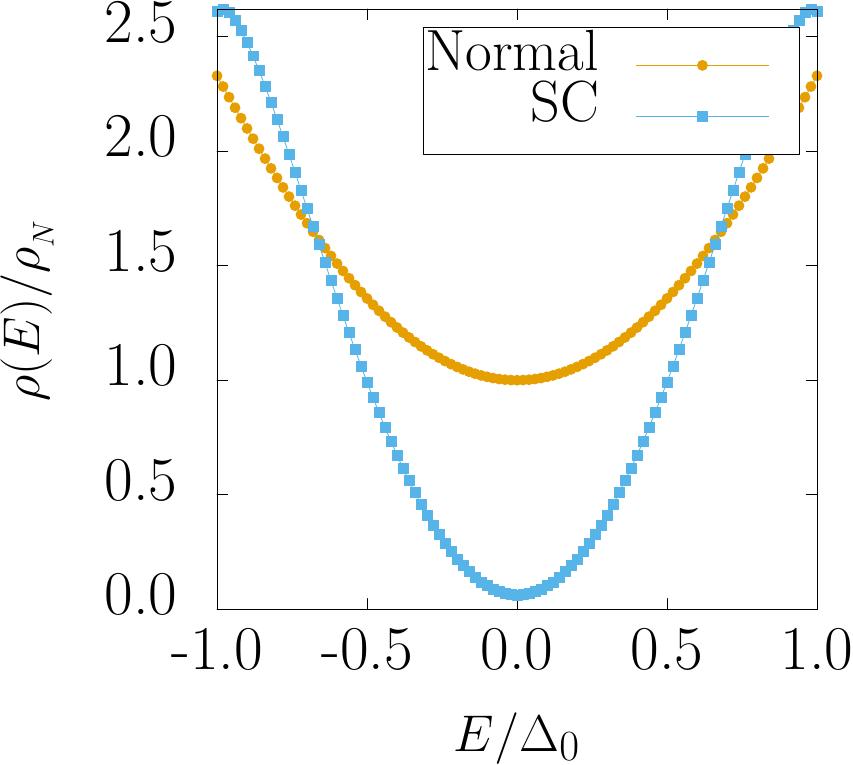}
}
\parbox{0.48\linewidth}{\centering (b)~$B_{1u}$
\includegraphics[width=\linewidth]{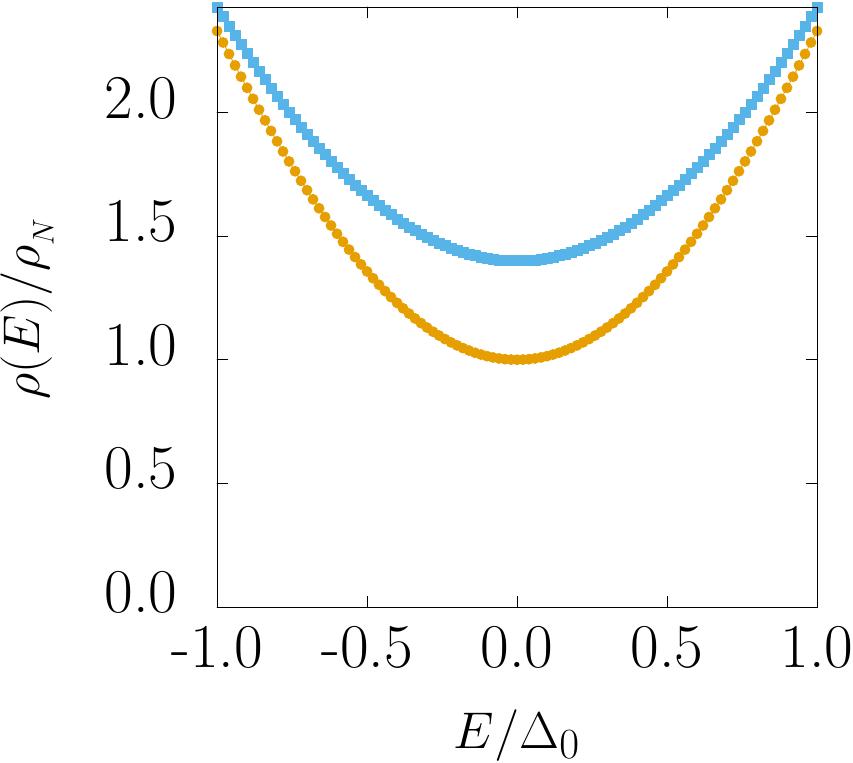}
}
\parbox{0.48\linewidth}{\centering (c)~$B_{2g}$
\includegraphics[width=\linewidth]{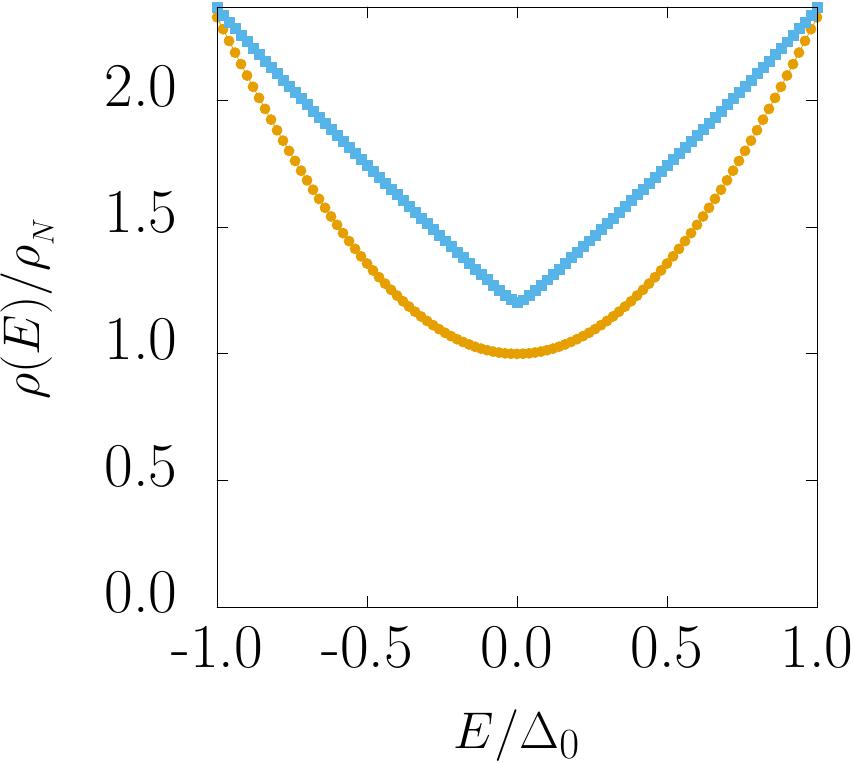}
}
\parbox{0.48\linewidth}{\centering (d)~$A_1$ 
\includegraphics[width=\linewidth]{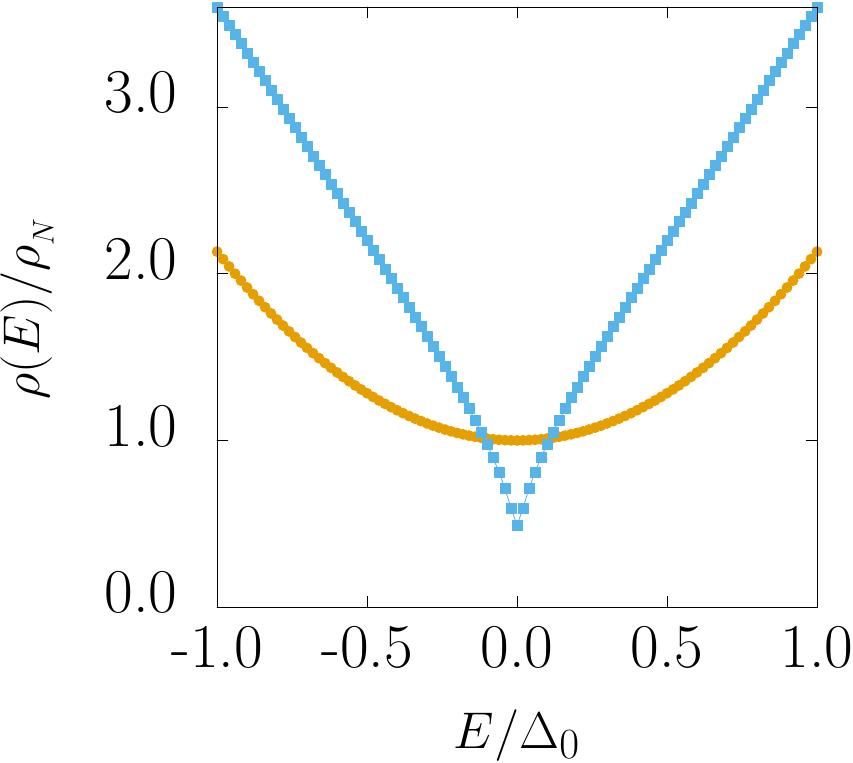}
}
\parbox{0.48\linewidth}{\centering (e)~$B_1$                   
\includegraphics[width=\linewidth]{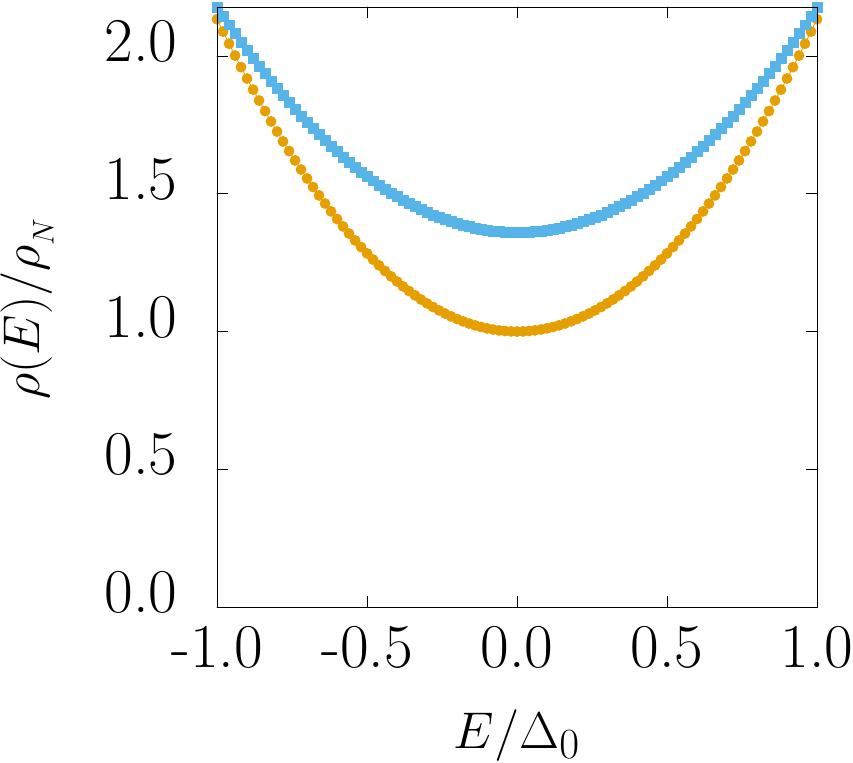}
}
\parbox{0.48\linewidth}{\centering (f)~$B_2$                   
\includegraphics[width=\linewidth]{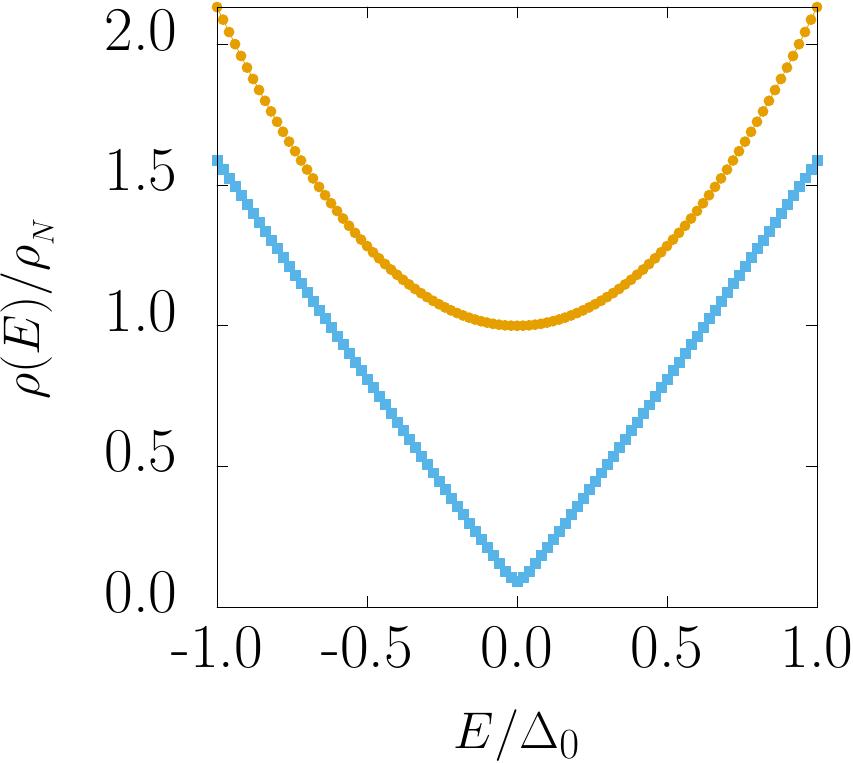}
}
\parbox{0.48\linewidth}{\centering (g)~$E_1$(triplet)
\includegraphics[width=\linewidth]{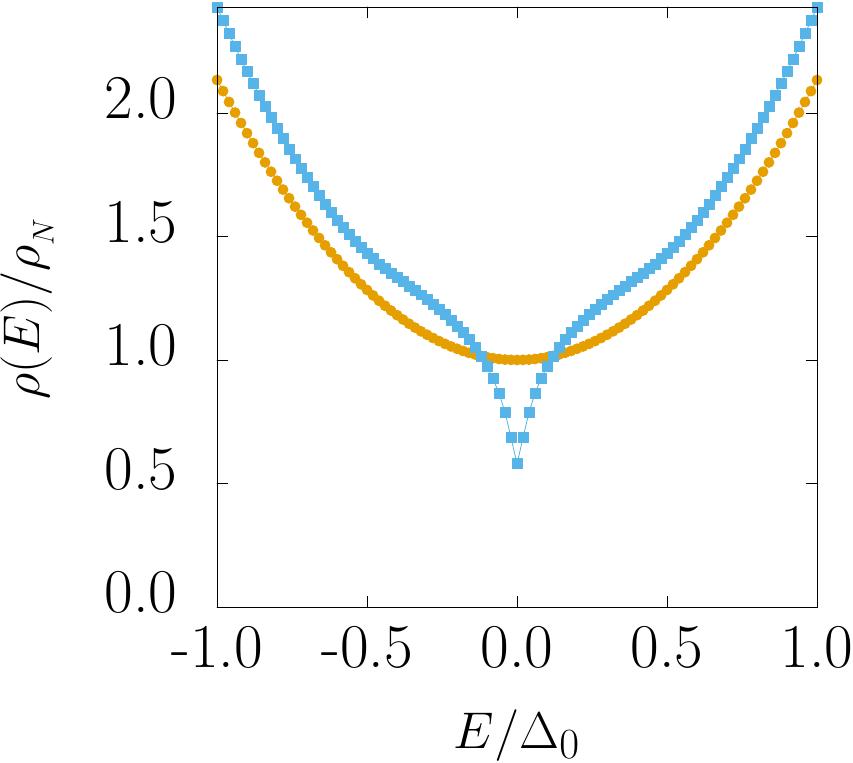}
}
\parbox{0.48\linewidth}{\centering (h)~$E_2$(singlet)
\includegraphics[width=\linewidth]{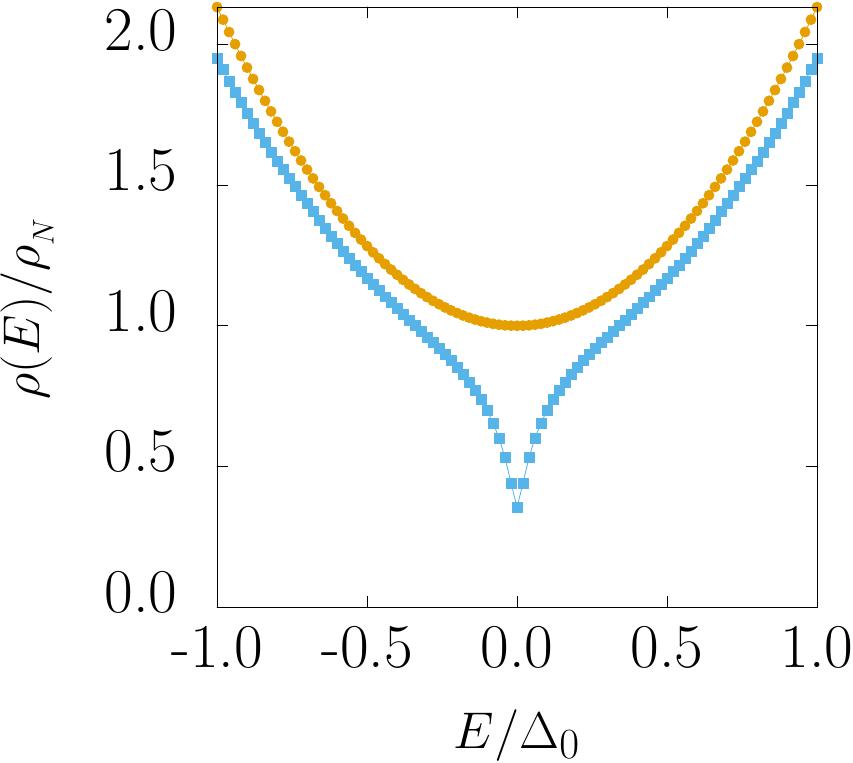}
}
\caption{Surface density of states on the (001) surface. The figures are shown in the same manner as Fig.~\ref{fig:SDOS_001}. The hopping parameters are set as the same values as in Fig.~\ref{fig:SDOS_001}. The amplitudes of the pair potential are set as $\Delta_0=0.2$.}
\label{fig:SDOS_001_app}
\end{figure}

\section{Numerical results of the surface density of states without zero energy peak}\label{sec:app}

\begin{figure}[htb]                                        
\parbox{0.48\linewidth}{\centering (a)~$A_{1g}$
\includegraphics[width=\linewidth]{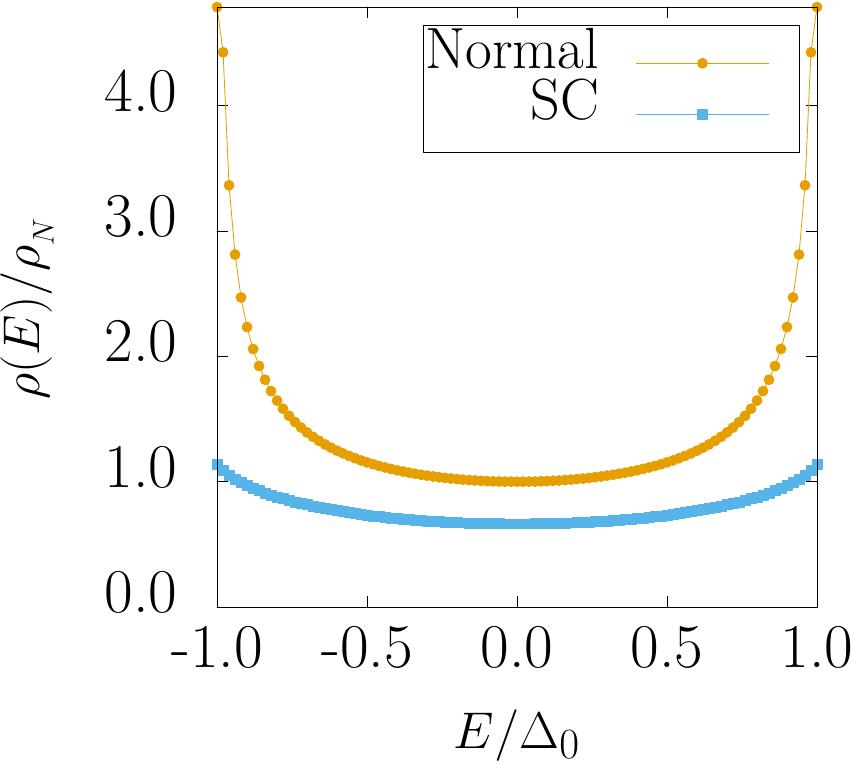}
}
\parbox{0.48\linewidth}{\centering (b)~$A_{2u}$
\includegraphics[width=\linewidth]{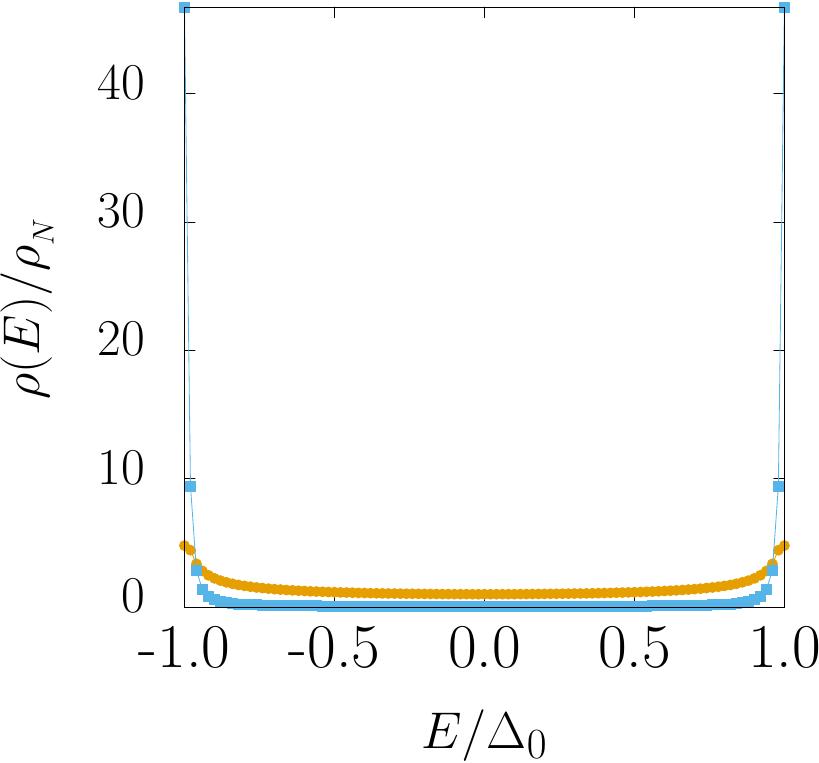}
}
\parbox{0.48\linewidth}{\centering (c)~$B_{1u}$
\includegraphics[width=\linewidth]{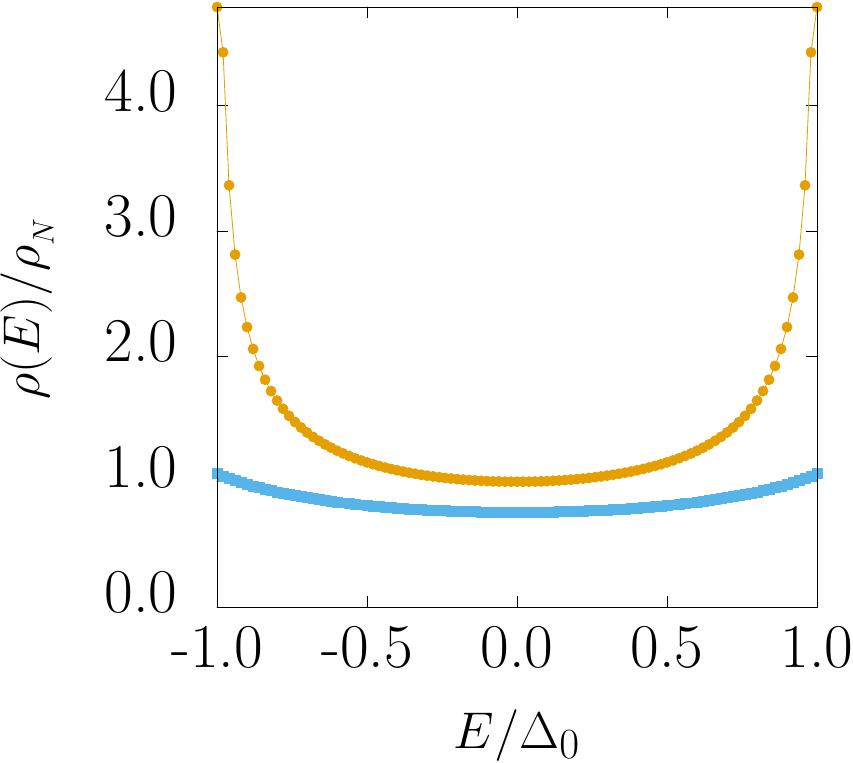}
}
\parbox{0.48\linewidth}{\centering (d)~$B_{2g}$
\includegraphics[width=\linewidth]{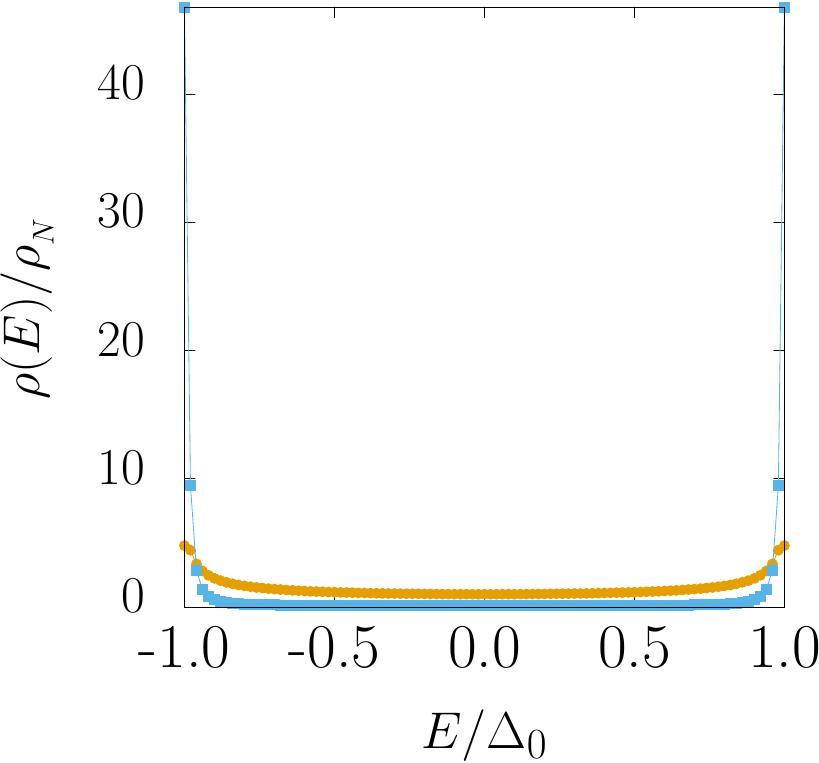}
}
\parbox{0.48\linewidth}{\centering (e)~$A_1$ 
\includegraphics[width=\linewidth]{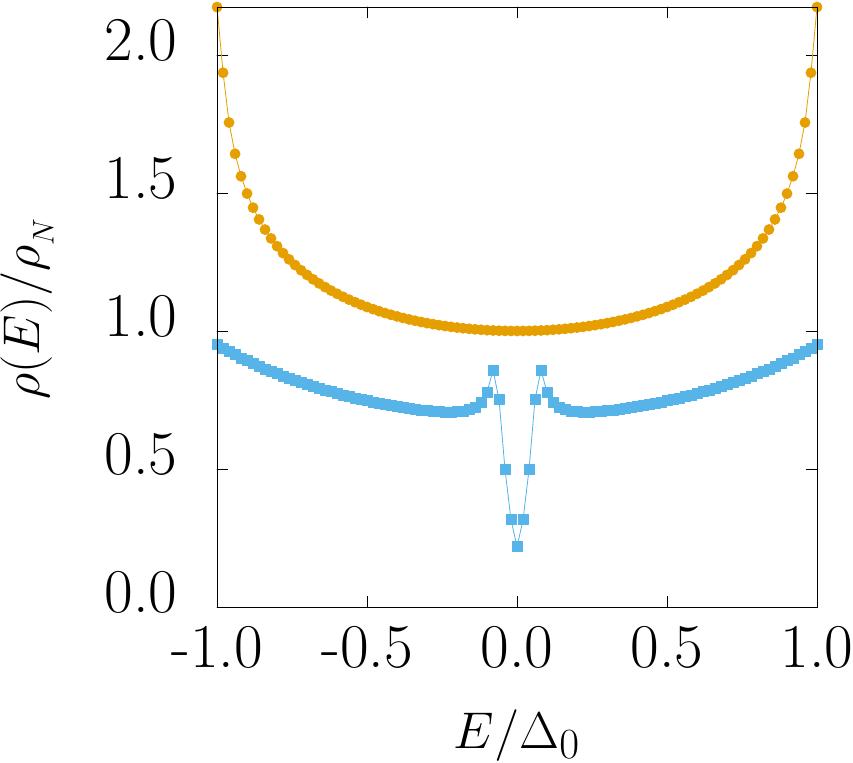}
}
\parbox{0.48\linewidth}{\centering (f)~$A_2$ 
\includegraphics[width=\linewidth]{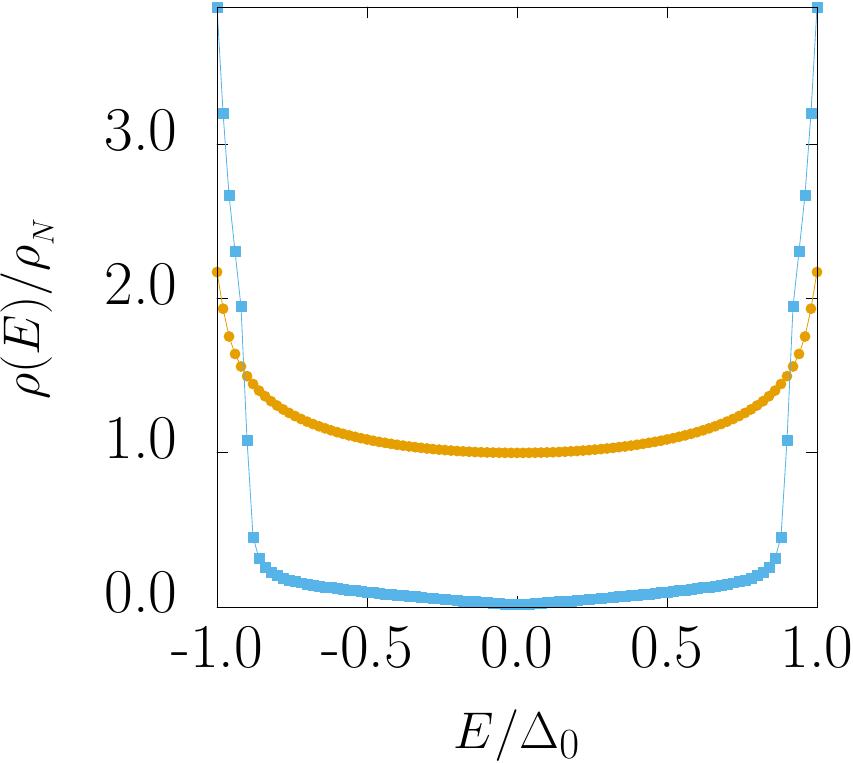}
}
\parbox{0.48\linewidth}{\centering (g)~$B_1$
\includegraphics[width=\linewidth]{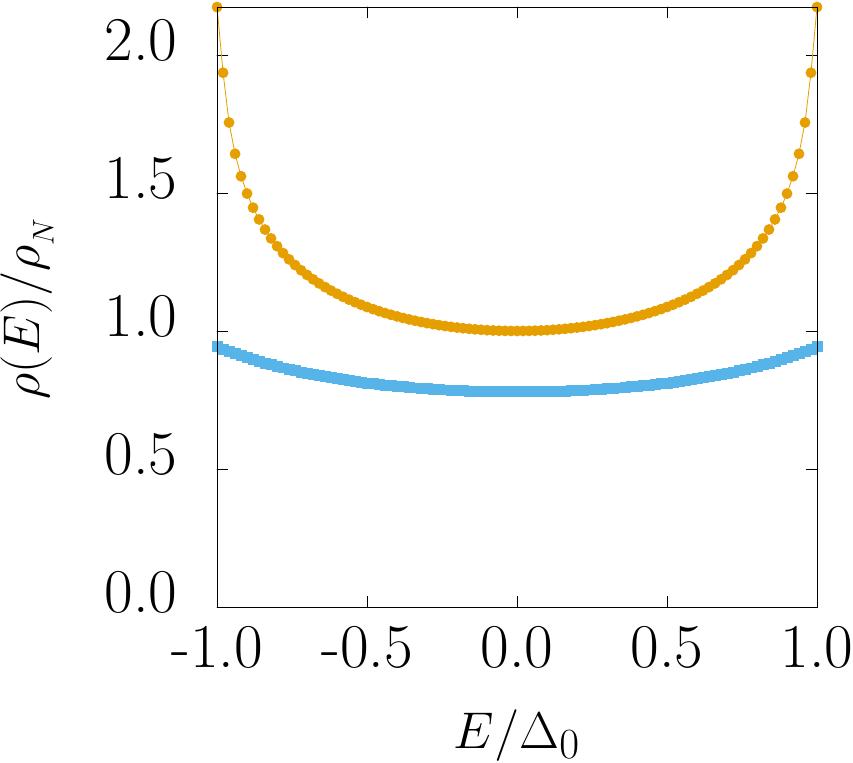}
}
\parbox{0.48\linewidth}{\centering (h)~$B_2$
\includegraphics[width=\linewidth]{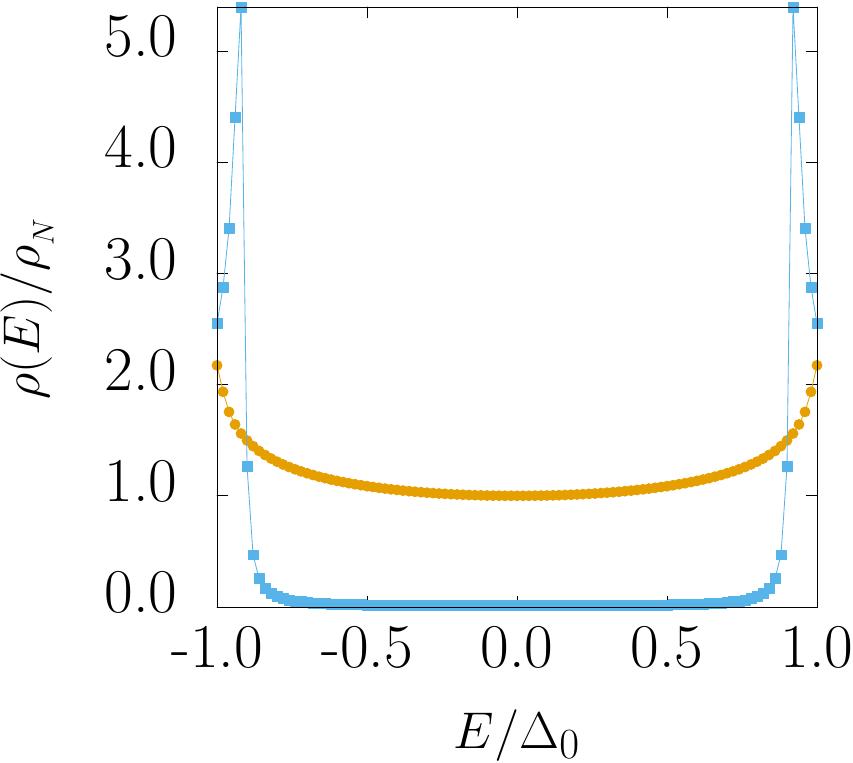}
}
\caption{Surface density of states on the zigzag surface. The figures are shown in the same manner as Fig.~\ref{fig:SDOS_001}. The hopping parameters and amplitudes of the pair potential are set as the same values as in Fig.~\ref{fig:SDOS_001_app}.}
\label{fig:SDOS_zigzag_app}
\end{figure}

Here, we show the SDOS for the other representations not exhibited in the main text. 
We show the SDOS on the (001) and zigzag surfaces in Figs.~\ref{fig:SDOS_001_app} and \ref{fig:SDOS_zigzag_app}, respectively.
The zero energy peaks do not appear for all the SDOS in Figs.\ref{fig:SDOS_001_app} and \ref{fig:SDOS_zigzag_app}.

\bibliographystyle{apsrev4-1}
\bibliography{HL.bib}

\end{document}